\documentclass[10pt,journal,compsoc]{IEEEtran}
\ifCLASSOPTIONcompsoc
  \usepackage[nocompress]{cite}
\else
  \usepackage{cite}
\fi
\ifCLASSINFOpdf
   \usepackage[pdftex]{graphicx}
\else
\fi
\usepackage[boxed,linesnumbered,ruled]{algorithm2e}
\makeatletter
\renewcommand{\@algocf@capt@plain}{above}% formerly {bottom}
\makeatother
\SetKw{KwBy}{step}

\usepackage{subcaption} % For subfigures
\usepackage{prettyref}
\newrefformat{eqn}{Equation~\ref{#1}}
\newrefformat{fig}{Figure~\ref{#1}}
\newrefformat{tab}{Table~\ref{#1}}
\newrefformat{sec}{Section~\ref{#1}}
\newrefformat{ssec}{Subsection~\ref{#1}}
\newrefformat{lst}{Listing~\ref{#1}}
\newrefformat{alg}{Algorithm~\ref{#1}}
\newrefformat{chapter}{Chapter~\ref{#1}}

\usepackage[htt]{hyphenat}

\usepackage{color}

\usepackage{multirow}

\begin{document}
%
% paper title
% Titles are generally capitalized except for words such as a, an, and, as,
% at, but, by, for, in, nor, of, on, or, the, to and up, which are usually
% not capitalized unless they are the first or last word of the title.
% Linebreaks \\ can be used within to get better formatting as desired.
% Do not put math or special symbols in the title.
\title{GPU-optimized Approaches to Molecular Docking-based Virtual Screening in Drug Discovery: A Comparative Analysis}
%
%
% author names and IEEE memberships
% note positions of commas and nonbreaking spaces ( ~ ) LaTeX will not break
% a structure at a ~ so this keeps an author's name from being broken across
% two lines.
% use \thanks{} to gain access to the first footnote area
% a separate \thanks must be used for each paragraph as LaTeX2e's \thanks
% was not built to handle multiple paragraphs
%
%
%\IEEEcompsocitemizethanks is a special \thanks that produces the bulleted
% lists the Computer Society journals use for "first footnote" author
% affiliations. Use \IEEEcompsocthanksitem which works much like \item
% for each affiliation group. When not in compsoc mode,
% \IEEEcompsocitemizethanks becomes like \thanks and
% \IEEEcompsocthanksitem becomes a line break with idention. This
% facilitates dual compilation, although admittedly the differences in the
% desired content of \author between the different types of papers makes a
% one-size-fits-all approach a daunting prospect. For instance, compsoc 
% journal papers have the author affiliations above the "Manuscript
% received ..."  text while in non-compsoc journals this is reversed. Sigh.

\author{Emanuele~Vitali,
        Federico Ficarelli,
        Mauro Bisson,
        Davide Gadioli,\\
        Massimiliano Fatica,
        Andrea R. Beccari,
        Gianluca Palermo, \IEEEmembership{Senior Member,~IEEE} 

\IEEEcompsocitemizethanks{\IEEEcompsocthanksitem E.Vitali, D. Gadioli and G.Palermo are with the Dipartimento di Elettronica, Informazione e Bioingegneria of Politecnico di Milano, Milan, Italy\protect\\
% note need leading \protect in front of \\ to get a newline within \thanks as
% \\ is fragile and will error, could use \hfil\break instead.
E-mail: name.surname@polimi.it

\IEEEcompsocthanksitem F.Ficarelli is with CINECA, Bologna and Dipartimento di Ingegneria dell'Energia Elettrica e dell'Informazione of Università di Bologna, Italy
% note need leading \protect in front of \\ to get a newline within \thanks as
% \\ is fragile and will error, could use \hfil\break instead.

\IEEEcompsocthanksitem M.Bisson and M.Fatica are with NVIDIA Corporation, Santa Clara, CA 95051, USA.
\IEEEcompsocthanksitem A.Beccari is with EXSCALATE, Dompé Farmaceutici S.p.A, Napoli..
}% <-this % stops an unwanted space
%\thanks{Manuscript received MONTH DAY, YEAR; revised MONTH DAY, YEAR.}
}

\IEEEtitleabstractindextext{%
\begin{abstract}
COVID-19 has shown the importance of having a fast response against pandemics. Finding a novel drug is a very long and complex procedure, and it is possible to accelerate the preliminary phases by using computer simulations. In particular, virtual screening is an in-silico phase that is needed to filter a large set of possible drug candidates to a manageable number.
This paper presents the implementations and a comparative analysis of two GPU-optimized implementations of a virtual screening algorithm targeting novel GPU architectures. 
The first adopts a traditional approach that spreads the computation required to evaluate a single molecule across the entire GPU. 
The second uses a batched approach that exploits the parallel architecture of the GPU to evaluate more molecules in parallel, without considering the latency to process a single molecule. 
The paper describes the advantages and disadvantages of the proposed solutions, highlighting implementation details that impact the performance.
Experimental results highlight the different performance of the two methods on several target molecule databases while running on NVIDIA A100 GPUs. The two implementations have a strong dependency with respect to the data to be processed. 
For both cases, the performance is improving while reducing the dimension of the target molecules (number of atoms and rotatable bonds). 
The two methods demonstrated a different behavior with respect to the size of the molecule database to be screened. While the latency one reaches sooner (with fewer molecules) the performance plateau in terms of throughput, the batched one requires a larger set of molecules. However, the performances after the initial transient period are much higher (up to 5x speed-up).
Finally, to check the efficiency of both implementations we deeply analyzed their workload characteristics using the instruction roof-line methodology.
\end{abstract}

% Note that keywords are not normally used for peerreview papers.
}

% make the title area
\maketitle

% To allow for easy dual compilation without having to reenter the
% abstract/keywords data, the \IEEEtitleabstractindextext text will
% not be used in maketitle, but will appear (i.e., to be "transported")
% here as \IEEEdisplaynontitleabstractindextext when the compsoc 
% or transmag modes are not selected <OR> if conference mode is selected 
% - because all conference papers position the abstract like regular
% papers do.
\IEEEdisplaynontitleabstractindextext
% \IEEEdisplaynontitleabstractindextext has no effect when using
% compsoc or transmag under a non-conference mode.

% For peer review papers, you can put extra information on the cover
% page as needed:
% \ifCLASSOPTIONpeerreview
% \begin{center} \bfseries EDICS Category: 3-BBND \end{center}
% \fi
%
% For peerreview papers, this IEEEtran command inserts a page break and
% creates the second title. It will be ignored for other modes.
\IEEEpeerreviewmaketitle

\IEEEraisesectionheading{\section{Introduction}\label{sec:introduction}}
% Computer Society journal (but not conference!) papers do something unusual
% with the very first section heading (almost always called "Introduction").
% They place it ABOVE the main text! IEEEtran.cls does not automatically do
% this for you, but you can achieve this effect with the provided
% \IEEEraisesectionheading{} command. Note the need to keep any \label that
% is to refer to the section immediately after \section in the above as
% \IEEEraisesectionheading puts \section within a raised box.

% The very first letter is a 2 line initial drop letter followed
% by the rest of the first word in caps (small caps for compsoc).
% 
% form to use if the first word consists of a single letter:
% \IEEEPARstart{A}{demo} file is ....
% 
% form to use if you need the single drop letter followed by
% normal text (unknown if ever used by the IEEE):
% \IEEEPARstart{A}{}demo file is ....
% 
% Some journals put the first two words in caps:
% \IEEEPARstart{T}{his demo} file is ....
% 
% Here we have the typical use of a "T" for an initial drop letter
% and "HIS" in caps to complete the first word.
Drug discovery is a long and costly process that aims at finding new drugs. 
Typically, this process involves several \textit{in silico}, \textit{in vitro} tasks  ( ranging from chemical design to toxicity analysis ) and \textit{in vivo} experiments.
Virtual screening is one of these tasks, which has to be performed at the beginning of the drug discovery process in the exploratory research phase.
This task aims at reducing the number of candidate drugs from billions of molecules to a number that can be managed with costly chemical experiments. 
Molecular docking represents but one stage of this step \cite{Lionta2014,Beccari2017NovelSP}.
It aims at estimating the three-dimensional pose of a given molecule, the \textit{ligand}, when it interacts with the target protein.
The \textit{ligand} is much smaller than the target protein, and we only consider a region of the protein, called \textit{pocket} (or binding site) in the docking procedure.
The \textit{pocket} is an active region of the protein where it is likely that an external small molecule can interact.
Molecular docking algorithm is in charge of positioning the \textit{ligand} inside the \textit{pocket} in the most suitable place. This means that it needs to perform a set of translation and rotation operations on the target \textit{ligand}.
Furthermore, it is possible to identify a subset of bonds -- named \textit{rotamers} or \textit{rotatable bonds} -- that split the \textit{ligand} into two disjoint fragments when they are removed.
These \textit{rotamers} can be rotated without changing the chemical properties of the \textit{ligand}.
Therefore, the algorithm must also consider the different shapes of the \textit{ligand} that can be generated from the rotation of all its \textit{rotamers}.

An efficient implementation of the virtual screening phase has two positive effects. On one side, it reduces the time to wait for the screening phase. On the other side, it permits the enlargement of the input chemical space, thus increasing the number of molecules to be evaluated. 
While these advantages are clear,  they are even more evident nowadays due to the COVID-19 pandemic. 
Indeed, when the pandemic started, several efforts have been kicked off around the world to find a therapeutic cure for the SARS-CoV-2 infection. Example in this direction are the COVID-19 HPC Consortium\footnote{https://covid19-hpc-consortium.org/} and the EXSCALATE4CoV project and league\footnote{https://www.exscalate4cov.eu/}.

The workload in the case of screening a large set of molecules is embarrassingly parallel since each ligand-pocket pair can be processed in parallel to the others.
This makes the use of large supercomputer infrastructure the most suitable target \cite{ph15010063,9250525} for urgent computation in the case of a pandemic given the possibility to have a simple data splitting across the nodes with lighter synchronization for I/O accesses \cite{9651263,9669317}.
Similar thoughts can be done when considering resources within the node. In particular, current supercomputers are mostly accelerated with multiple GPU cards, and the workload can be further split for each card.

In this paper, we analyze two different GPU implementations of a high-throughput \textit{in silico} virtual screening application, LiGen \cite{beccari2013ligen}, to compare their behavior given the different parallelization strategies.

Both of these implementations target NVIDIA GPU and are written in CUDA. However, they have an orthogonal approach where the first implementation is a synchronous, latency-oriented one, while the second is an asynchronous implementation that uses a batched approach.
In the first implementation, called \textit{latency implementation} from now on, we exploit the GPU parallelism to shorten the computation time required to dock a single ligand by evaluating the different poses and different atoms in parallel. 
This is the classic approach for the acceleration of molecular docking applications, used in AutoDock \cite{SantosMartins2021AcceleratingAW}, and in previous versions of LiGen \cite{vitali2019exploiting}.
In the second implementation, we approach the problem of parallelizing the computation from a different perspective: we exploit the GPU parallelism to evaluate several ligands in parallel, and a single ligand is always evaluated by a single warp.  
The warp is a collection of threads, 32 in current implementations, that are executed simultaneously by a Streaming Multiprocessor, SM, therefore this is considered the basic unit of execution on a GPU.
For this reason, we will define this version as \textit{batched implementation}.

The remainder of the document is organized as follows: \prettyref{sec:related} briefly describes the state of the art in the field and related approaches applied to virtual screening.
\prettyref{sec:target} briefly describes the target architecture with the CUDA-related abstraction. \prettyref{sec:application_descr} describes the target application and the two different implementations under analysis. 
\prettyref{sec:experimental_res} reports the experimental results obtained by the two implementations highlighting the performance characteristics and limitations, together with a deep profiling analysis on the use of the resources.
Finally, \prettyref{sec:conclusion} concludes the paper.

\section{State of the Art}
\label{sec:related}
High throughput virtual screening has been widely applied in the latest years during the early stage of drug discovery. 
Indeed, this helped in finding some novel drugs \cite{doi:10.1021/jm0341913,doi:10.1517/17460441.3.8.841,MACCONNACHIE1999369}.
There are several steps that are required to perform a virtual screening campaign \cite{PMID:26094053}; however, in this work, we will focus on the molecular docking step.

Many pieces of software have been created during the latest years to this end, both open sources\cite{dock, autodock} and commercial\cite{glide, flexx}.
There are two main approaches to this task, where the first one is to use a deterministic approach while the second favors a random-based approach.
Random-based approaches use well-known techniques to create different poses of a \textit{ligand} and measure their interactions with the protein \textit{pocket}.
Examples of these are MolDock \cite{thomsen2006moldock} and Gold \cite{jones1997development} where genetic algorithms are used or Glide  \cite{glide} and MCDock \cite{liu1999mcdock} where the technique used are the Monte Carlo simulations.
However, this approach has a big drawback since its results may not be reproducible in their entirety. 
This may be a blocking issue for some pharmaceutical companies that refuse to start the expensive in-vitro and in-vivo phases without a reproducible result.
For this reason, sometimes a deterministic approach is required. 
Examples of deterministic approaches are BIGGER \cite{palma2000bigger}, DOCK \cite{dock}, LiGen \cite{beccari2013ligen}, and Flexx \cite{flexx}. These approaches use deterministic algorithms that are able to modify the shape of the ligands by leveraging their torsional bonds. 

Many molecular docking applications were born as single workstation applications; however, given the amount of complex elaboration that has to be performed, they quickly evolved into High-Performance Computing (HPC) applications.
As we can see from this survey \cite{survey}, different techniques were studied to improve the capabilities of this software and scale them to HPC machines. 
Among them, there are obvious approaches such as scaling with MPI \cite{zhang2013message} to more complex solutions such as developing ad-hoc scripts to wrap the main kernel and deploy it to different nodes with different data \cite{zhang2008dovis}.
In the latest years, we have seen the rise of heterogeneous clusters in HPC, where next to the CPU, several GPUs are used as accelerators. 
This can be seen by looking at the TOP500 \cite{top500}, where many of the listed machines adopt this paradigm.
For this reason, some of these molecular docking applications have been modified in order to be able to exploit these co-processors \cite{medusadock,sukhwani2009gpu,Korb2011,Fang2016,Sanchez-Linares2012,tang2022accelerating}.
In particular, MedusaDock \cite{medusadock} achieves a 1.54X overall speedup, and GeauxDock \cite{Fang2016} has a 3.5X speedup thanks to the GPU porting.
Other application show a better behavior on the GPU and have double digits speedup, such as PIPER \cite{sukhwani2009gpu} with a 17X speedup, AutoDock-Vina \cite{tang2022accelerating} with a 50X speedup and PLANT \cite{Korb2011}, that reported a 60X speedup.
%Not all the offloading have been this succesful, since there is still a strong dependency from the algorithm. 
%An example is \cite{medusadock}, where the application is accelerated with the usage of GPU (NVIDIA P100) and achieves a 1.54× overall speedup compared to CPU-only execution.
%Other examples can be found in literature, such as \cite{tang2022accelerating}, where a speedup in the range of 21x to 50x is reached on AutoDock-Vina, or 
%\cite{sukhwani2009gpu}, where a speedup of 17x is reached on PIPER.
%
The latest GPU porting of AutoDock \cite{legrand2020, SantosMartins2021AcceleratingAW,9407259} has been optimized for running on the Summit supercomputer \cite{glaser2021} to support COVID-19 related researches. %, where the usage of the GPU achieves a 50x speedup over the single-threaded CPU application using a TITAN V GPU. %
%This application has been used in  to perform a large virtual screening campaign during the COVID-19 pandemic on the Summit supercomputer.

%In this paper, we are going to analyze two different implementations of LiGen that target NVIDIA GPUs.
In this paper, we focus only on the GPU porting of the LiGen application. We describe and analyze two different parallelization approaches considering the peculiarities of the target workload and GPU devices.
%LiGen was born as a CPU-only MPI application, that used a provider-consumer paradigm to distribute the workload across different cores and different nodes of a supercomputer \cite{gadioli2021tunable}.
%Then, we decided to offload the main computation kernels to GPU using OpenACC as first attempt \cite{vitali2019exploiting}, and then CUDA for having the largest virtual screening campaign ever run during the first wave of the COVID-19 pandemic \cite{EXSCALATE22TETC}.
%
LiGen is an MPI application that distributes the workload across different nodes of a supercomputer \cite{gadioli2021tunable}, and it has been used for the largest virtual screening campaign ever run (>70 billion ligands and 12 viral proteins) during the first wave of the COVID-19 pandemic \cite{EXSCALATE22TETC}. 
%Then, we decided to offload the main computation kernels to GPU using OpenACC as first attempt \cite{vitali2019exploiting}, and then CUDA for having the largest virtual screening campaign ever run during the first wave of the COVID-19 pandemic \cite{EXSCALATE22TETC}.
%Then we decided to offload the main computation kernels to GPU. In the first implementation, we used OpenACC to offload the kernels \cite{pbio}, and we also studied a way to balance the CPU and the GPU workload \cite{vitali2019exploiting}. 
%
%In this paper, we focus only on the GPU porting of the LigGen application. We describe and analyze two different parallelization approaches used to optimize the application, considering the peculiarities of the target workload and GPU devices.
%This paper focuses only on the GPU porting by describing and analysing two different parallelization approaches for molecular docking-based virtual screening problem, while considering the peculiarities of the target workload and GPU devices. 
%This paper focuses only on the GPU porting by describing and analysing two different optimized implementations of the molecular docking-based virtual screening application considering the peculiarities of the target workload and GPU devices. 

%Finally, we decided to use CUDA to optimize the usage of the GPU, and we developed the two versions that we are going to analyze in this paper.

\section{Target Architecture}
\label{sec:target}
We target NVIDIA GPUs, and we use the CUDA language to exploit the maximum potential of the architecture.
In this section, we will provide some general considerations on these architectures that have been used as guidelines when writing both implementations.

\begin{figure}[h!]
\centering
\includegraphics[width=\columnwidth]{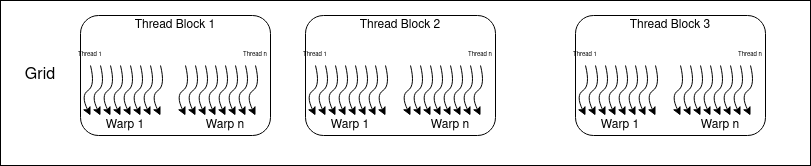}
\caption{GPU Threads Layout}
\label{fig:gpu_par}
\end{figure}

The first one is that more levels of parallelism are needed to fully exploit the GPU. 
We must aim at a SIMT (Single Instruction Multiple Threads) approach, where different threads are executing the same operation on different inputs. 
This must be done since threads are the inner parallelism level in the CUDA hierarchy. 
They are organized in blocks (of max 1024 threads), which are themselves organized in grids.
A block is mapped on a single streaming multiprocessor (SM), while the grid is distributed across the SMs on the GPU. 
This organization is visible in \prettyref{fig:gpu_par}

An important detail in organizing the code is the attention to warps:
a warp is a unit of thread scheduling in the GPU and consists of 32 threads. A warp executes one common instruction at a time, so full efficiency is achieved when all threads in the warp execute the same instruction. If the execution paths of threads in a warp diverge via a conditional branch, the warp executes each path disabling the threads that are not on that path (inactive threads). That introduces an overhead since the instructions on the two paths are not executed in parallel.
The most common instructions that can cause warp divergence are conditional instructions, such as loops, if statements, and so on.
Since warp divergence can negatively impact the performance of a code, it is essential to minimize the number and (more importantly) the length of divergent execution paths within the warps as much as possible.

\begin{figure}[h!]
\centering
\includegraphics[width=\columnwidth]{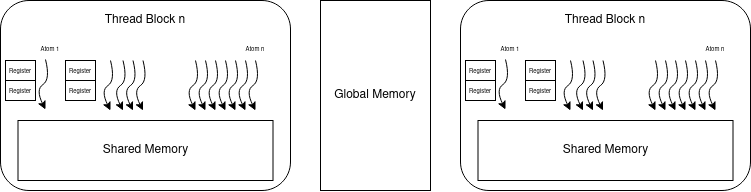}
\caption{GPU Memory Layout}
\label{fig:gpu_mem}
\end{figure}

Another important feature of the GPU that must be considered is the memory hierarchy, visible in \prettyref{fig:gpu_mem}. 
It is really important that all the data are close to where they are needed, however, GPUs have a small cache memory and a different organization with respect to the CPU.
There are 3 different levels of memory on the board:
\begin{itemize}
    \item Global Memory: the slowest (2000 GB/s of bandwidth on A100) and largest memory (up to 80GB on A100) available on the accelerator. 
    This is the memory to which data from the host can be directly copied.
    \item Shared Memory: a small, low-latency, and high-bandwidth
    memory shared among the threads of a block, and accessible by all of them.
    Threads of a block can load shared memory with data from global memory and, by using thread synchronization, shared memory can be used as a scratchpad to implement user-defined data caches. SMs cannot run more blocks than the amount of shared memory they are equipped with allows, and thus shared memory usage of a kernel can limit occupancy.
    \item Register set: the fastest memory space available to threads. Each SM has 65536 32-bit registers that are used partitioned among the threads of the blocks running on it. In addition to the amount of shared memory required by each block, also the number of registers per thread that a compiled kernel requires determines how many blocks can concurrently run on an SM and, if too high, can limit the kernel occupancy.

\end{itemize}

\section{Application Description}
\label{sec:application_descr}

\textit{LiGen}\cite{beccari2013ligen} is an extreme-scale molecular docking application, designed to run on High Performance Computers. 
\textit{LiGen} uses a mixed approach for docking a \textit{ligand} in the target \textit{pocket}.
It starts considering geometric features, then it estimates the actual physical and chemical interaction for the most promising \textit{ligand} poses.
In the Docking algorithm, we focus only on the geometrical docking phase, used to filter out incompatible \textit{ligands}.
This problem is embarrassingly parallel, since the evaluation of every \textit{ligand} is independent of the others, and we do not need to communicate anything.
We use this characteristic of the problem to address multi-node parallelism.
This allows us to distribute the ligands across the different nodes and cores of the supercomputer, with a very small communication overhead (we do not need to synchronize the nodes during the evaluation of a ligand, but only while reading from the input or writing the output.
In this paper, we will explore two strategies to address the grid and warp levels of the GPU parallelism using the characteristics of the docking algorithm.

\begin{algorithm}[t]

\KwIn{Target Pocket and the initial pose of the ligand}
\KwOut{The score of the evaluated poses}

\Repeat{$Pose\_id < N$}{

$Generate\_Starting\_Pose(Pose\_id);$

\For{$angle\_x\gets0$ \KwTo $360$ \KwBy $12$}
{
    $Rotate (angle\_x, Pose\_id);$
    
    \For{$angle\_y\gets0$ \KwTo $360$ \KwBy $12$}
    {
        $Rotate (angle\_y, Pose\_id);$
        
        $Evaluate\_Score(Pose\_id)$        
    }
}
  
\For{$fragment$ in $ligand\_fragments$}{
    \For{$angle\gets0$ \KwTo $360$ \KwBy $36$}
        {
$Rotate (fragment, angle,pose\_id)$

$Bump\ Check (fragment,pose\_id)$

$Score (fragment,pose\_id)$
        }
     }
}
$Select\ Poses()$

\Repeat{$Pose\_id < K$}{

   $Rescore\ Pose(Pose\ Id)$ 
   
}

\caption{Pseudo-code of the original algorithm that performs the docking for the CPU.}
\label{alg:one}
\end{algorithm}

Due to the high number of degrees of freedom (rotations, translations, and \textit{rotamer} induced changes of shape), it is unfeasible to perform an exhaustive exploration of the possible poses of a \textit{ligand}.
For this reason, in \textit{LiGen} we use a greedy optimization heuristic with multiple restarts.

\prettyref{alg:one} shows the pseudo-code of the docking procedure of a single \textit{ligand}.
The outer loop generates \textit{N} different initial poses for the target \textit{ligand}, increasing the probability to avoid local minima.
Each iteration of the outer loop aims at docking the $i^{th}$ initial pose of the \textit{ligand}.

Within the body of the outer loop, the docking algorithm is divided into two sections.
The first one (lines 3-9) performs rigid rotations of the \textit{ligand}, to find the best alignment with the target \textit{pocket}, according to the scoring function.
We will refer to this section of the algorithm as Rigid Rotation or Alignment.
In the last section of the algorithm (lines 10-16), we optimize the shape of the \textit{ligand} by evaluating each \textit{fragment} in an independent fashion (line 10).
In particular, we rotate each \textit{fragment} to find the angle that maximizes the scoring function without overlapping with the other atoms of the ligand (lines 11-15).
We will refer to this section of the algorithm as Optimize Pose.
%we might change the \textit{ligand} structure in an unpredictable way, since some of the rotations may not happen in the same order invalidating the outcome of the application.
Once all the poses have been docked, we have a pose filtering step, that selects the optimal poses according to the geometrical scoring function, also discarding the poses that are too similar to each other (line 18).
Once the pose filtering phase is completed, we have the final loop (lines 19-21) where the most promising poses are re-scored according to the chemical scoring function.

Finally,
let us highlight data dependencies:
different iterations of the outer loop are independent since every initial pose represents the actual starting point of the docking algorithm.
In the alignment phase, we perform rigid rotations to select the most suitable alignment of the different initial poses of the \textit{ligand} for the target \textit{pocket}. This operation can be parallelized, since we are only interested in the best outcome, and all the atoms are moved in every rotation.
%The first step aligns the initial pose of the \textit{ligand} with the \textit{pocket} using rigid rotations.
However, in the Pose Optimization phase, need to evaluate each \textit{fragment} sequentially, since some of the atoms that compose a \textit{fragment} may be included in another \textit{fragment} and the evaluation is stateful.
Therefore, if we parallelize the pose optimization over the \textit{fragments}, we would destroy the molecule structure.
On the other hand, evaluating all the possible angles in a combinatorial fashion would be too expensive to compute and it would change the algorithm. %, since it will increase the number of evaluated poses by a large number.
Nonetheless, it is still possible to parallelize the inner loop performing the rotations, the validity check, and the scoring of the optimized poses.
In the Rescore function, we need to evaluate all the ligand's atoms against all the pocket's atoms, but since no atom is going to be moved, all these comparisons can be independent.
However, to evaluate the score, we need to accumulate the ligand-pocket interactions over each of the ligand's atoms. 
This complicates the approach, which needs to have some reductions and synchronization points.

\subsection{Latency Implementation}

The first implementation that we are going to analyze is the \textit{latency implementation}.

\begin{figure}
\centering
\includegraphics[width=0.8\columnwidth]{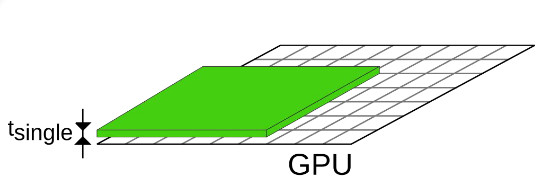}
\caption{Latency Implementation GPU Usage}
\label{fig:latency}
\end{figure}

The idea behind this approach is to keep a synchronous interface, where a single \textit{ligand} is docked in every call to the dock function.
This approach is the same as the previous implementation of LiGen \cite{vitali2019exploiting} and allows us to focus only on the acceleration of the most expensive kernels without having to modify the whole application approach.
We try to distribute on the GPU as much as possible the operation that we have to perform, trying to make it as parallel (and fast) as possible to execute. 
%In this way we can target the whole GPU grid and have a set of small and quick kernels.
In this way, we try to use all the resources of the GPU to perform each of the kernels implementing the docking operations.
These kernels have a very small execution time and aim at freeing the hardware resource for other kernels, as reported in \prettyref{fig:latency}.
This solution follows the paradigm of accelerating as much as possible the compute-intensive kernels, which is the traditional approach in accelerating an application.

With this idea in mind, having seen the algorithm in \prettyref{alg:one}, we can map the intermediate level (multiple grids) on the multiple restarts where every restart has a different starting pose.
All the poses are independent and can be evaluated in parallel. 
Indeed, at this level, all the data can be considered independent and do not need synchronization before the final score evaluation phase, where the best-scored pose is selected.
The inner level of parallelism (threads) is given by the atoms: every \textit{ligand} is composed of several atoms that are moved according to some rotations and translation. 
This is the SIMT approach that we are searching for since the same instruction (rotate or translate) is applied to different data (the atoms).

On the host side, we rely on a multi-threaded approach to instantiate several different kernels (on different streams) in order to always have the GPU working at full capacity.
Every \textit{ligand} will be tied to a host thread, that is tied to an asynchronous queue (CUDA stream) and a reserved space in the GPU memory.
The reservation of the space at thread level instead of at ligand level allows us to allocate and deallocate that memory only once in the lifetime of the thread. This is a first optimization that allows saving a lot of memory operations, since this memory space is not linked to the docking of a single ligand, but is linked to the lifetime of the application.
The drawback of this approach is that we need to allocate the \textit{worst case} space, and this must be known at compile time. This introduces a limitation on the maximum size of the processed ligands. However, this is not a real issue for the application since it can be changed at compile time.
Moreover, some data structures (such as the pocket space) can be shared among all the threads that are using the same GPU: this can be done since they are read-only data structures, not modified in the docking process. 
The access to the pocket does not follow a coalesced pattern but the access point is given by the x,y, and z coordinates of the atom and, for this reason, has a random pattern.
Random accesses in memory are a costly operation in GPU since they disable the coalesced access mechanism that allows providing data to all the threads in a warp with a single read operation.
However, there is a feature in CUDA that allows for improving the performance in these situations, which is the texture cache.
Texture caches allow organizing data in 2D or 3D spaces and are optimized for semantic data locality. 
This means that accessing points in the space that are close to the previous ones is usually faster since they should already be cached. 
We expect that rotations and translations in the 3D space (\prettyref{alg:one}) will not place atoms "too far" across the different iterations.
For this reason, we use the texture cache to store the protein pocket values.

On the other hand, when multi-dimensional arrays are needed and they have to be accessed from different thread-blocks, it is very important to organize the data in a way that allows the reads to be coalesced. 
For this reason, we extensively use CUDA pitched arrays in storing temporary values that are needed across kernels.
%Pitched arrays are an instrument provided by CUDA that automatically inserts padding at the end of every line of multi-dimensional arrays, to optimize memory accesses. In particular, it avoids bank-conflicts and allows coalesced accesses.
Pitched multi-dimensional arrays are an instrument provided by CUDA and are allocated with rows padded to a size that ensures that each row starts at an address that meets the alignment requirements for coalescing.

In the following part of this section, we are going to analyze some of the main kernels to highlight the design choices of this approach.

\subsubsection{Alignment}
This kernel covers lines 3-9 of \prettyref{alg:one}. 
As we already mentioned, in this kernel, all the different poses may be tested in parallel, performing a reduction at the end of the kernel to select the best pose.
We decided to distribute the computation as follows: every grid is going to perform the alignment on a different starting pose, and a single thread is going to perform the rotation on a different rotation matrix, with different x, and y angles combinations.
Since we need the different poses only to evaluate the score, we can avoid storing them thus calculating the score on the fly.
In this way, we distribute as much as possible the computation across the GPU, speeding up the execution of the kernel. 

\subsubsection{Pose Optimization}
This kernel covers lines 10-16 of \prettyref{alg:one}.
The loop starting at line 10 cannot be parallelized since the rotations of the fragments are sequential. 
However, the inner loop starting at line 11 instead can be accelerated with the GPU, since once again we only need to retrieve the best-scored angle.
This loop needs to test a smaller amount of poses, but the operations to perform on each angle are more, since we also need to perform the bump\_check function.
For this reason, we have assigned an entire warp (32 threads) to every angle, and we have different threads cooperating in the scoring of the pose obtained.
We still assign the grid level parallelism to the different initial poses, to maximize the utilization of the accelerator.

\subsubsection{Rescore Pose}
This kernel covers the last loop in \prettyref{alg:one} (lines 19-21).
This kernel
%we need to
compares every ligand's atom against all the pocket atoms to evaluate some characteristics, and we need to accumulate some of those results.
The kernels are organized to occupy the whole GPU by mapping every different pose to a different thread block, while the threads are used to evaluate the $p \times l$ couple of atoms characteristics.
\subsection{Batched Implementation}
\label{sec:batched}
The second version of the application is the \textit{batched implementation}.
This implementation follows a completely different paradigm than the \textit{latency} one. Instead of
using the whole GPU to process a single ligand at a time, we pack it with as many ligands as
possible that are processed in parallel (using fewer resources per ligand).
This approach follows a different paradigm, similar to the one described in \cite{5529656}, that is one of the benchmark of the NAS \cite{saini1996parallel} benchmark suite, used to estimate the upper achievable limits of floating-point performances on a system since it requires almost no communication to process the data.
This approach is possible since the amount of data per ligand is quite limited (up to 20KB input - 1MB output).

\begin{figure}
\centering
\includegraphics[width=0.8\columnwidth]{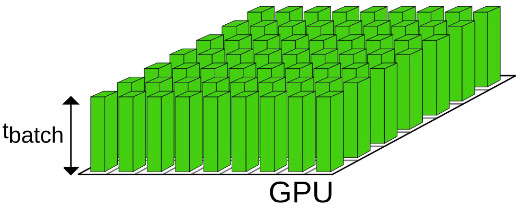}
\caption{Batch Implementation GPU Usage}
\label{fig:batch}
\end{figure}

With this approach, the time to process a single ligand $t_{batch}$ will be greater than the time required by the \textit{latency} implementation $t_{latency}$; however, many more ligands will be processed in parallel during the time $t_{batch}$ as shown in \prettyref{fig:batch}. As long as the size of the batch of ligands processed in parallel is greater than $t_{batch}/t_{latency}$, this implementation is expected to deliver higher throughput than the \textit{latency} one.

This implementation forces us to modify the whole approach of the application since we need to first load several ligands in a single batch and then launch the processing kernels when the batch is full. 
This mandates a paradigm switch, leading us to adopt an asynchronous paradigm where different CPU threads push ligands in the batch and another CPU thread is in charge of launching the kernels when the batch is full.

In this implementation, the external parallelism (CUDA grid) is addressed by docking multiple ligands at the same time, while the thread-level parallelism is addressed by distributing across a single warp the set of operations to perform on the atoms of a single ligand.
There are some criticalities that we need to address for this approach to be successful.

The obvious one is that we need a large number of ligands to fully utilize the GPU. This is not a concern since, as we mentioned in \prettyref{sec:introduction}, we target a design space with millions or billions of candidate molecules.

The second one is that since we are processing batches of ligands concurrently, the overall kernel time will be dictated by the slowest warp of the grid, i.e. the warp assigned to the ligand that requires more operations. For this reason, we need to balance the size of the ligands that are collected in a single batch.
This is also important to make efficient use of registers and shared memory, two very precious and scarce resources in the GPU. To make the batched kernels run as fast as possible,
ligand data that are used often, like atoms coordinates and fragments indices, are kept in registers and shared memory to access them more efficiently. Since this requires defining at compile time the resources used by the kernels, balancing the sizes of the ligands in the batches allows maximizing the usage of those statically allocated resources.

For this reason, we have clustered the ligands in 5 different batches, according to their number of atoms: (0, 32], (32, 64], (64, 96], (96, 128], (128, 160].
The number of ligands that are accumulated in each batch before being processed by the GPU, depends on the maximum number of atoms in its range. 
For each range, we used kernels compiled to reserve a precise number of registers per thread such that each warp can hold at most a number of atoms equal to the upper limit of the range. Thus the size of each batch is set equal to the maximum number of warps that can be concurrently active on all GPU's SMs with the respective kernel. We determined this number by using the \textit{cudaOccupancyMaxActiveBlocksPerMultiprocessor} function.

\begin{figure}
\centering
\includegraphics[width=\columnwidth]{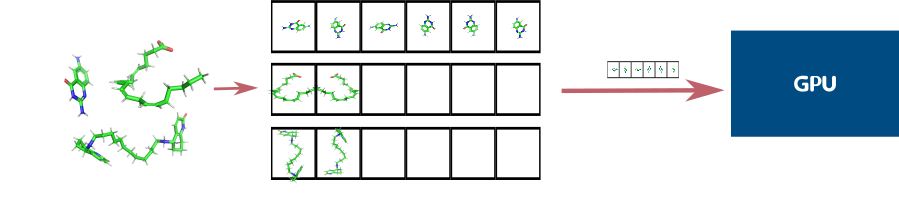}
\caption{Graphical representation of the batch creation process: all incoming ligands are divided into batches according to their characteristics, and only when a bucket is full is sent to the GPU.}
\label{fig:bucketizer}
\end{figure}

Moreover, as can be easily imagined by looking at \prettyref{alg:one}, the loop over the number of fragments in the Optimize Pose phase of the algorithm can cause a strong imbalance if the number of fragments in the ligands belonging to the same batch varies significantly.
For this reason, we also need to cluster the \textit{ligands} by their number of fragments.
We decided to group them by four (i.e. ligands with 0-3 fragments are clustered in one batch, ligands with 4-7 in another, and so on). This decision is a compromise between having the ligands as similar as possible and avoiding the explosion of the number of different batches.
Considering all of these divisions, we have a matrix of buckets where we collect ligands with similar features. 
This aims at reducing the disparity between the ligands that need to be processed in a batch, to improve the efficiency of the computations. A graphical representation of this process is provided in \prettyref{fig:bucketizer}

The kernels developed for this implementation declare the array parameters as {\em const \_\_restrict\_\_} so that the compiler can automatically use cached loads for them. Moreover, since we read them only once to copy their content in the register/shared memory we use regular allocations instead of pitched ones.

In the following part of this section, we are going to analyze the main kernels' organization.

\subsubsection{Alignment}

Even if there are no dependencies between the poses, in this implementation, we are limiting the resources used to evaluate a single ligand.
For this reason, the implementation follows a sequential approach, closer to the original algorithm.
Indeed, we evaluate the "step rotation matrices", on both axis, and we store them in the constant memory.
Then we perform the two nested loops, using a warp to distribute the atoms of the ligand (i.e thread 0 will rotate atom 0, 32, 64, ...).
After the rotation, we will score the pose (by using a warp sum reduction) and store the new pose, using the register set.
During the different iterations, we are going to keep track of the best score and of the pose that generated it.
At the end of the loop, we are going to perform a single rotation to put the ligand in the best pose.

\subsubsection{Pose Optimization}
In this case, the intrinsic sequentiality of the fragments' evaluations is not a concern. 
Indeed, since we only have one warp to perform the whole computation, we use the warp parallelism only for the rotation, bump evaluation, and scoring functions.
In particular, for each fragment, we load the mask in the shared memory.
Then for each angle, we evaluate the rotation matrix and perform the rotation. After that, we check the internal bumps using the whole warp (up to 5 atoms per thread), and we compute the score only if the pose does not present any internal bump. 
Even in this circumstance, the kernel is closer to the original algorithm and its CPU implementation.

\subsubsection{Rescore Pose}
Even for the final kernel, we only use a single warp per ligand to perform the rescoring operations. 
In particular, in this kernel, we have two sequential loops: the first one on the poses, and the second on the ligand atoms. 
The warp parallelism is used to perform the inner loop, where every atom of the ligand is coupled with the atoms of the protein and the chemical properties are evaluated.
In this kernel, the shared memory is used to store some read-only dictionaries with the chemical properties of the atoms plus the atoms' distance matrix, while the coordinates of the atoms are read directly from the global memory and stored in registers when needed.

\section{Experimental Results}
\label{sec:experimental_res}

In this section, we are going to compare the two implementations in terms of throughput on different datasets and different conditions.

In particular, we are interested in two types of analysis: 
the first one consists of the comparison of the behavior on the preprocessed dataset, where we try to cluster the ligands according to their characteristics in terms of the number of atoms and number of fragments, and it is carried out in  \prettyref{ssec:preprocessed_ds},.
Since we have seen from \prettyref{sec:application_descr} that the algorithm is sensitive to these two data features, we want to analyze the two implementations to see if there is a common behavior or if they change according to the dataset.
Then we are going to compare the speedup on these sets of clustered datasets.
The second analysis regards the scaling of the throughput of the application according to the size of the dataset, and is done in \prettyref{ssec:scaling}: 
we want to know if one of the implementations is always optimal, or (and this is the expected behavior) if it depends from the dataset size. 
In this second circumstance, in particular, we are interested in finding what is the size of the dataset that triggers the optimality change. 
This analysis will be performed on a preprocessed dataset and on a real-world dataset where no preprocessing is done so the atoms' size and fragments are not known \textit{a priori}.

The third experiment, reported in \prettyref{ssec:realdb}, will show the performance of both implementations on a real-world public dataset, taken from the mediate initiative\cite{mediate}.

Finally, in \prettyref{ssec:workload}, we report an in-depth analysis of the workload done with the \textit{instruction roofline methodology}. This analysis is done to see the different resource utilization with different input categories (small, medium, and large ligands).

\subsection{Experimental Setup}
To perform the docking experiment, we target a machine that resembles an HPC node, equipped with 2 CPU AMD Epyc 7282 2.80GHz 16 core and one NVIDIA A100 GPU, connected with PCI-E 4.0. 

\subsection{Preprocessed Datasets}
\label{ssec:preprocessed_ds}
The first set of experiments wants to show the throughput of the two implementations when we are running at the best of the application capabilities (i.e. after the initial setup phase).
We have docked several datasets of 50K ligands each with different characteristics, where the smallest dataset consists of ligands with 20 Heavy Atoms and 1 fragment, and the most difficult one is composed of ligands with 50 heavy atoms and 20 fragments. 
In this context, we define as a heavy atom every non-hydrogen that composes the molecule.
We need to point out that having the same number of heavy atoms does not mean that all the \textit{ligands} belong to the same batch since Ligen will group them according to the total number of atoms, which also comprehends the hydrogens.

\begin{figure}[t]
	\centering
	\begin{subfigure}{0.95\columnwidth}
		\includegraphics[width=\columnwidth]{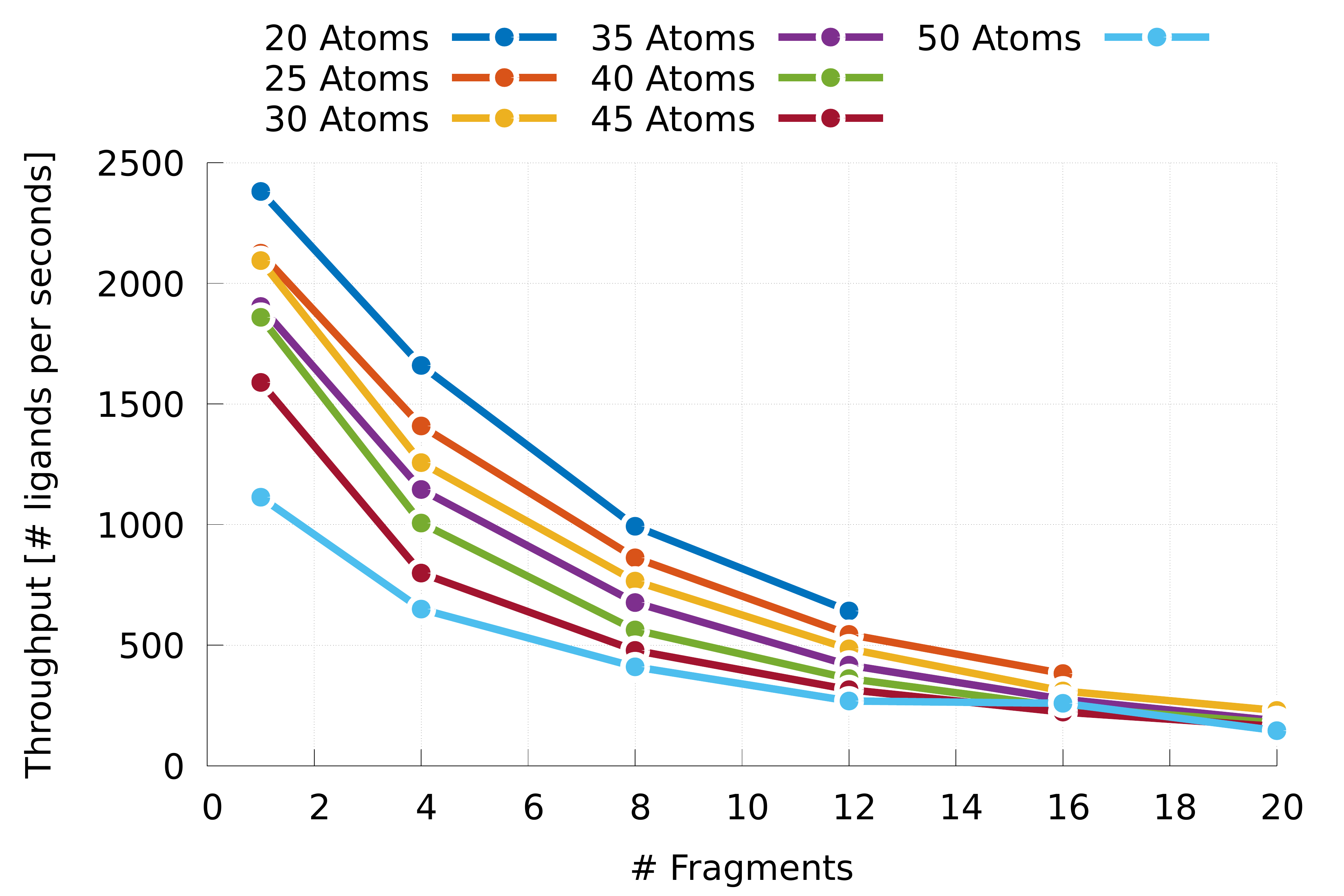}
		\caption{Latency}
		\label{fig:atmlat}
	\end{subfigure}
	\begin{subfigure}{0.95\columnwidth}
		\includegraphics[width=\columnwidth]{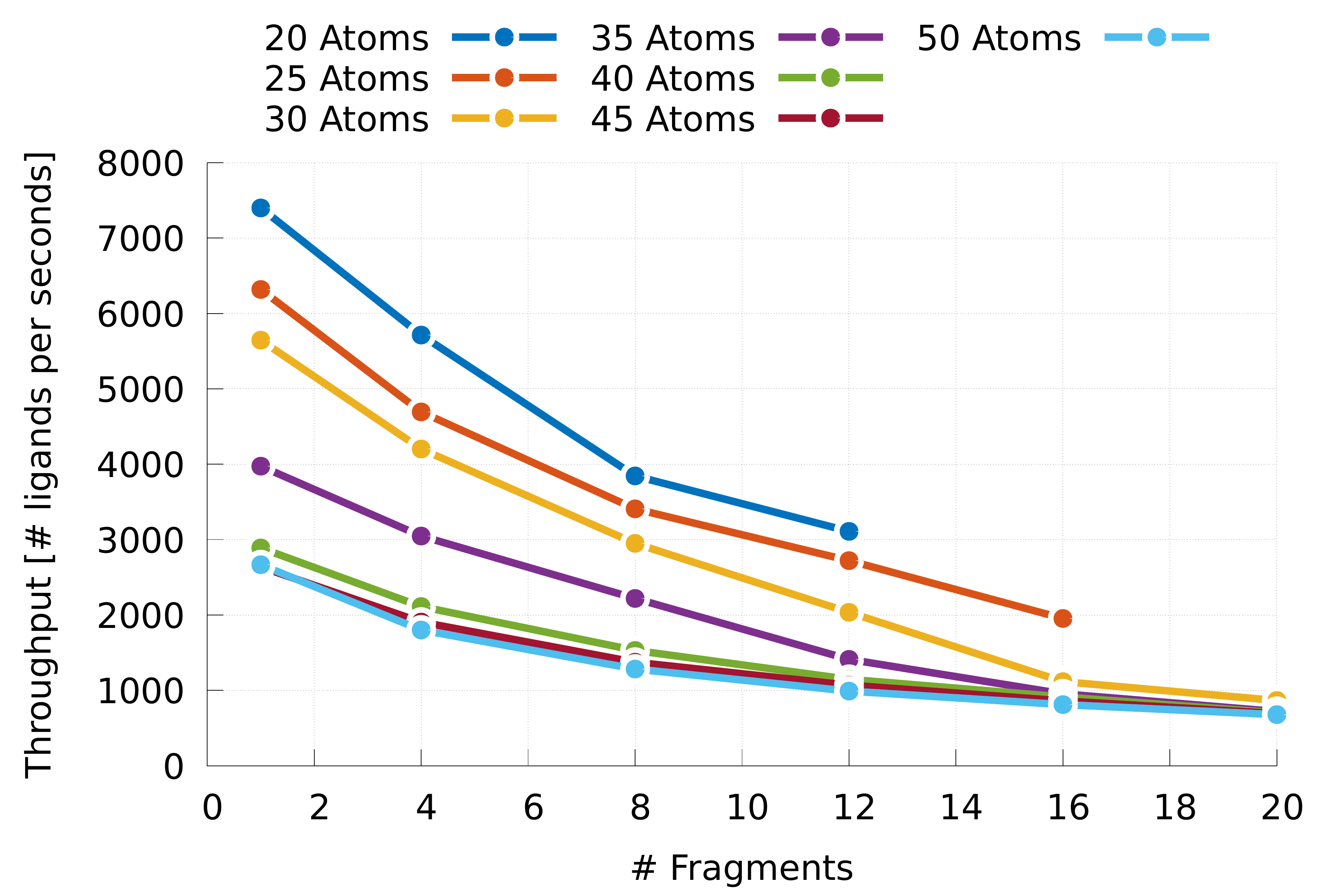}
		\caption{Batched}
		\label{fig:atmbat}
	\end{subfigure}
	\caption{Throughput of the two implementations with the different datasets, organized by the number of atoms and increasing the number of fragments on the X axis.}
	\label{fig:atm}
\end{figure}

We plot in \prettyref{fig:atm} the varying of the throughput according to the change in the number of fragments (x-axis). We can see that this data feature heavily impacts the throughput. The two implementations show similar behavior, going from a high throughput value with 1 fragment and slowing down always more with the increase of the number of fragments.
However, if we look at the y-axis, we can notice that the batched implementation is way faster than the latency one, on average by 3 times.

\begin{figure}[t]
	\centering
	\begin{subfigure}{0.95\columnwidth}
		\includegraphics[width=\columnwidth]{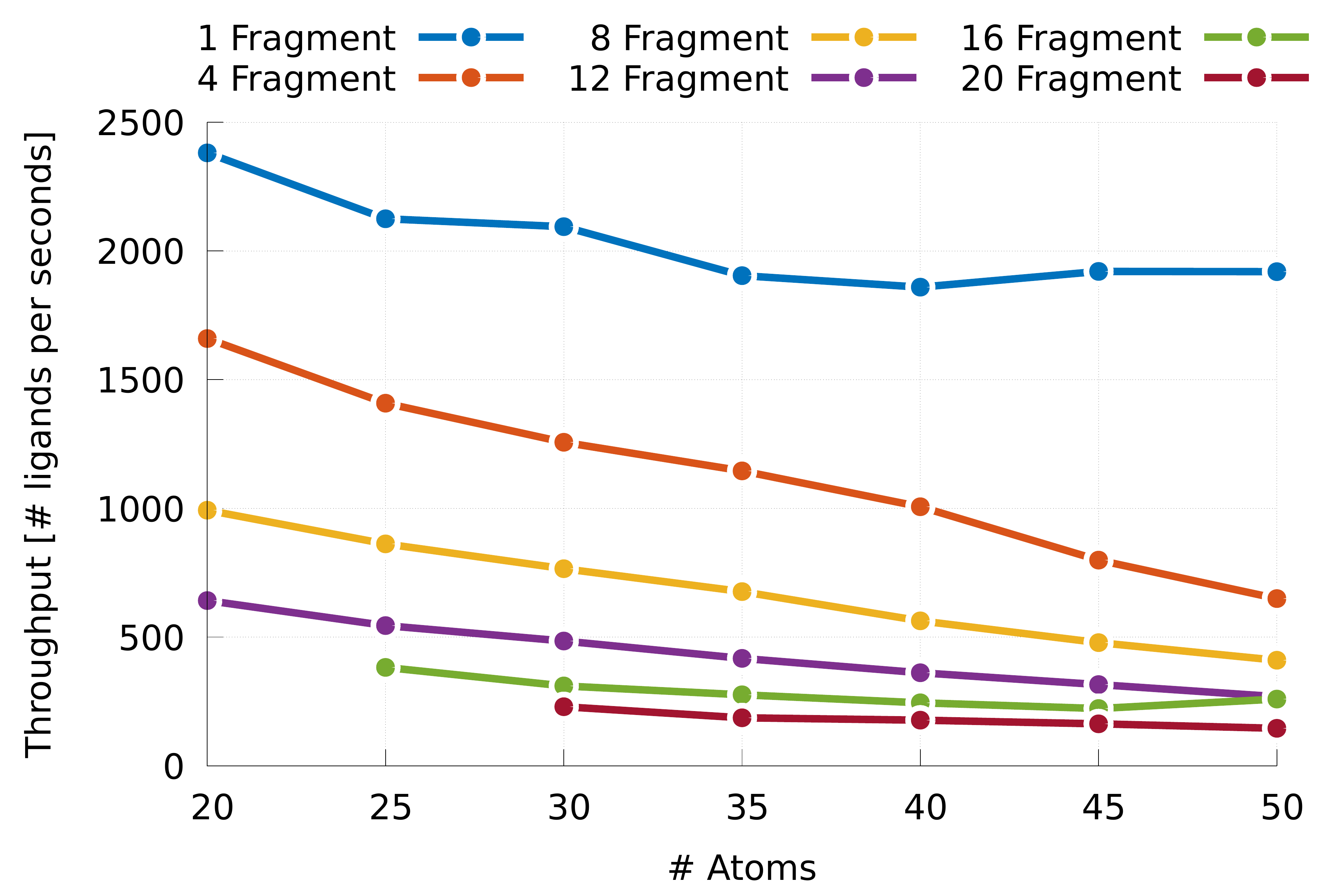}
		\caption{Latency}
		\label{fig:fraglat}
	\end{subfigure}
	\begin{subfigure}{0.95\columnwidth}
		\includegraphics[width=\columnwidth]{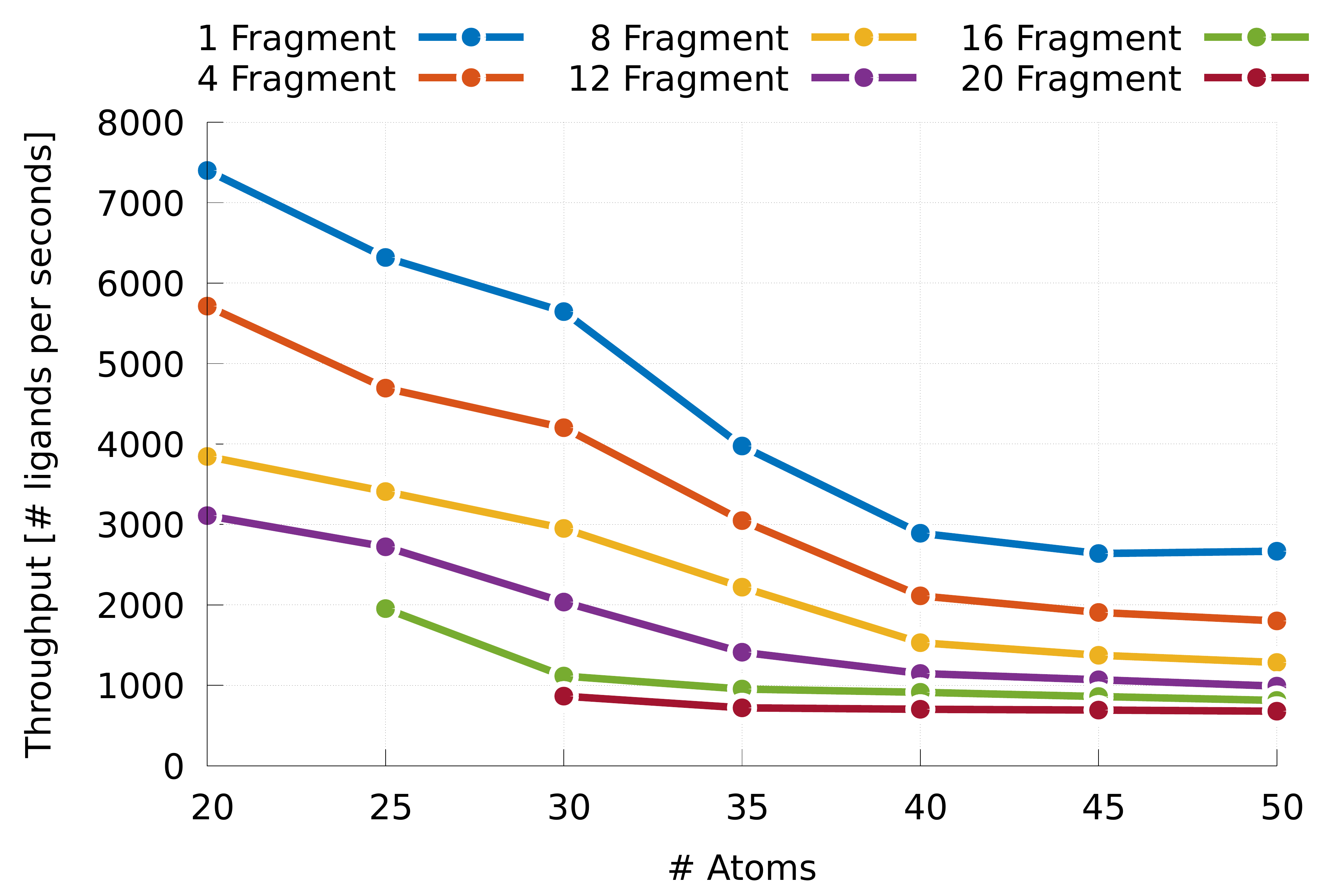}
		\caption{Batched}
		\label{fig:fragbat}
	\end{subfigure}
	\caption{Throughput of the two implementations with the different datasets, organized by the number of fragments and increasing the number of atoms on the X axis.}
	\label{fig:frag}
\end{figure}

In \prettyref{fig:frag} we report the varying in the throughput at the change of the number of atoms (plotted on the x-axis). We can notice that in this case, the behavior is a little different. 
The latency implementation has a smaller throughput degradation if we change the number of atoms with a constant amount of fragments, while the batched implementation has a more marked throughput loss.
However, since it starts from a higher throughput it still performs better than the latency implementation, in the worst case by 1.37x

\begin{figure}
\centering
\includegraphics[width=\columnwidth]{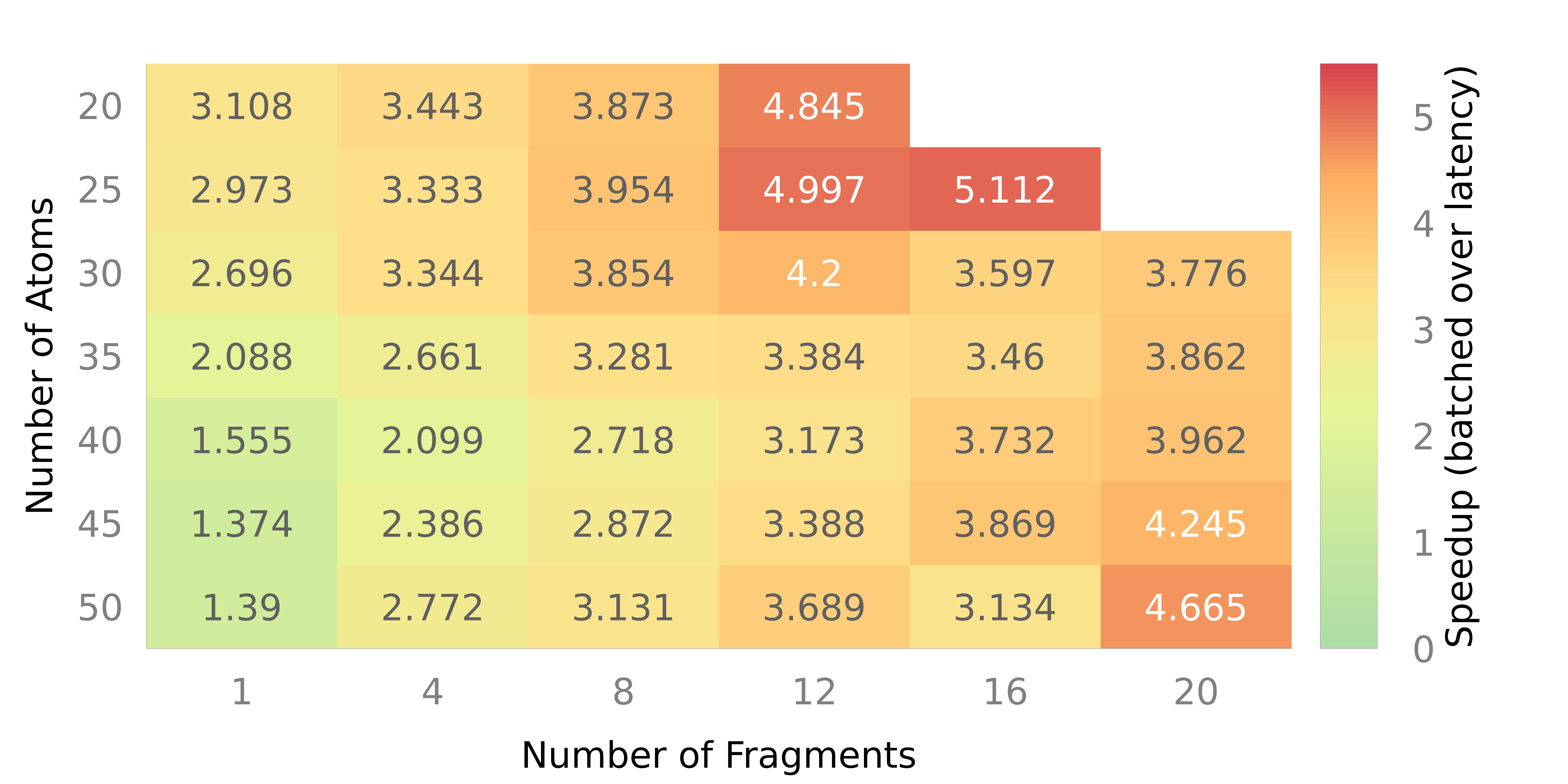}
\caption{Speedup Heatmap of the batched version against the latency one for the different homogeneous datasets of 50K ligands with the same characteristics.}
%\textbf{number atoms total}}
\label{fig:speedup}
\end{figure}

To conclude this analysis, we can see in \prettyref{fig:speedup} the heatmap of the speedup obtained by the batched implementation compared to the latency implementation, with several datasets of 50K ligands.
As we can see, the batched implementation is always better than the latency one, given this dataset dimension on a single GPU.
However, we can notice that the amount of speedup changes according to the characteristics of the \textit{ligands}: the batched implementation behaves dramatically better with a lower number of atoms and with a higher number of fragments.% dramatically increases on the most complex ligands, while for the simplest ones (i.e. with a low number of fragments) the speedup is lower.

\subsection{Scaling Analysis}
\label{ssec:scaling}
\begin{figure}
\centering
\includegraphics[width=\columnwidth]{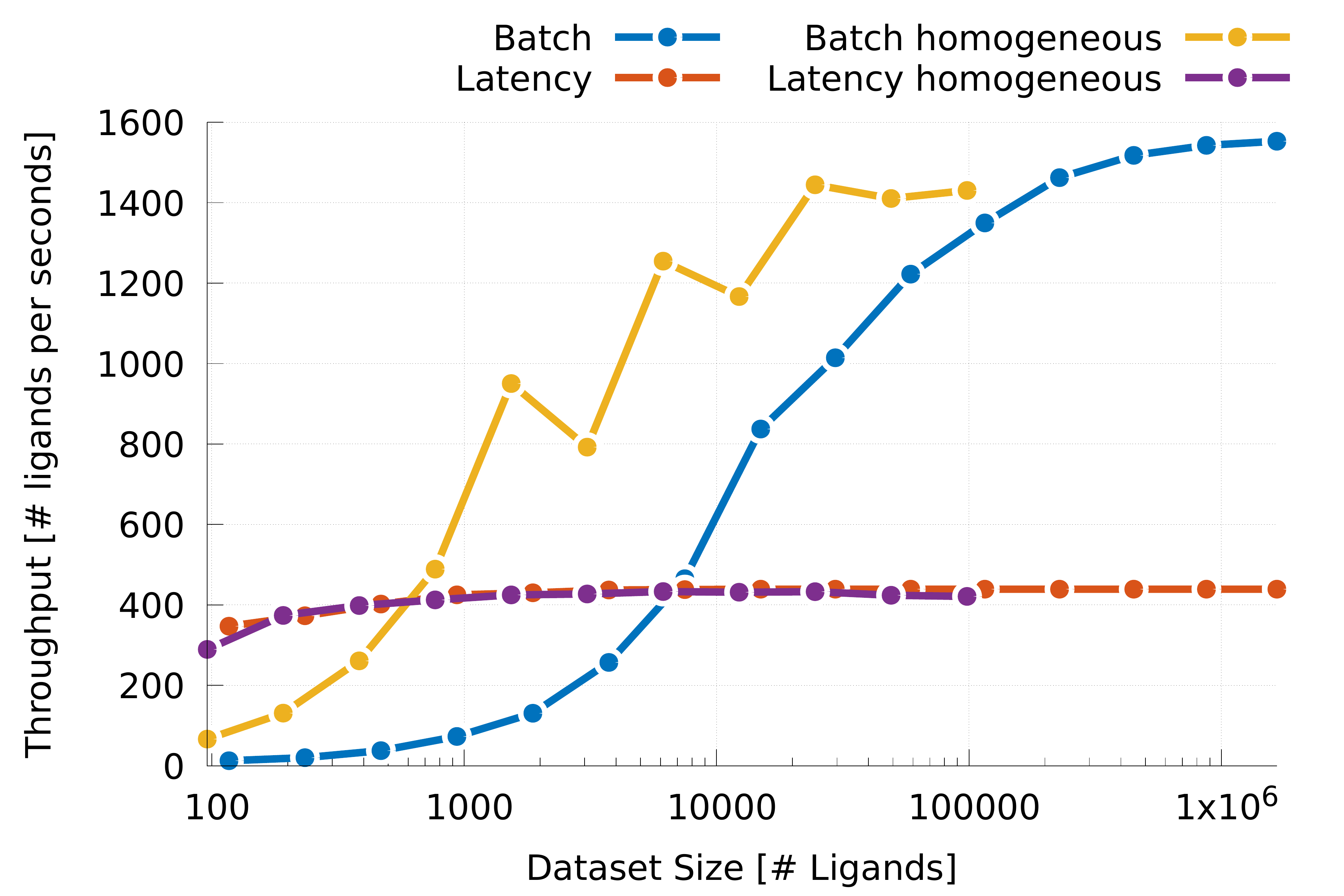}
\caption{Scaling analysis}
\label{fig:scaling}
\end{figure}

With this experiment, we aim at finding what is the minimal database size to reach throughput optimality with both implementations, and we are interested in seeing what is the impact of the dataset composition on this size.
\prettyref{fig:scaling} reports the growth of the throughput (y-axis) at the varying of the dataset size (x-axis).
As we can see, with small datasets the latency implementation outperforms the batched implementation. 
This happens because the batched implementation waits until the batch size is reached and distributes the computation on different CUDA warps. If the dataset is too small and does not reach the size of the batch, we are going to underutilize the GPU, and this explains why in these circumstances the latency implementation performs better. 
However, after a certain threshold, we can see that the batched implementation overtakes the latency implementation (with almost exponential growth) until it reaches its saturation point (with a total speedup of around 3.5x).
This behavior is observed in both the homogeneous dataset (purple and yellow lines) and the heterogeneous one (blue and red lines). 
The only difference between the two is when the batched implementation overtakes the latency one, and this happens for the homogeneous dataset one order of magnitude faster.
It is interesting to notice the fluctuations of the throughput in the yellow line (homogeneous batched implementation).
We did use for this analysis a preprocessed dataset where we grouped molecules with 35 Heavy Atoms and 12 Fragments. 
As we mentioned previously, the batches are created according to the total number of atoms. for this reason, we can notice that some \textit{ligands} are processed in different batches.
This explains the loss of performance that happens sometimes when we increase the dataset size: some ligands are added that have a different number of atoms and are inserted in a different bucket, thus forcing the application to process a small batch which we know is not optimal.
Finally, we can also notice that the growth phase of the batching application when using a mixed dataset ends at almost $10^6$ ligands.
This means that to get the maximum out of this implementation, we need to dock a very large dataset with at least $10^6$ ligands for each GPU involved in the computation. 
On the other hand, for the homogenous dataset 20K ligands are enough to stabilize the throughput.

\subsection{Real World Datasets}
\label{ssec:realdb}
Finally, we want to evaluate the applications on a real-world dataset. 
This dataset comes from the MEDIATE \cite{mediate} consortium.
This dataset contains several files where ligands are clustered according to their characteristics and every file contains a different category of ligands: Commercial, Natural, Drugs, and Peptides.

The largest sets are the files that are collected in the "Commercial" category. In particular, they are clustered in three files where molecules are selected according to their molecular weight (MW),for a total amount of ligands of around 5 million.
The first one contains ligands with MW lower than 330, the second set has ligands with MW between 330 and 500, the last contains all the ligands with MW higher than 500.

The Drugs category contains known drugs, including the set of safe-in-man drugs, commercialized or under active development in clinical phases. The total amount of known drugs is around 10 thousand.

The Natural category contains two sets of molecules: Foods and Natural Products. 
They are taken from the FooDB online database \cite{foodb}. FooDB is the world’s largest and most comprehensive resource on food constituents, chemistry, and biology. 
It provides information on many of the constituents that give foods their flavor, color, taste, texture, and aroma.
We have more than 200 thousand natural products, while the foods are around 65 thousand ligands.

Finally, peptides were generated by mixing in a combinatorial way all 20 natural amino acids. They are collected in three files according to the number of amino acids that compose the peptide. In particular, 2AA contains dipeptides (peptides formed by two amino acids), 3AA contains tripeptides and 4AA contains tetrapeptides.
All peptides have been constructed with an extended structure and have been optimized with MOPAC 2016 \cite{mopac}. 
They have been protected with acetylation of the N-terminal end and the addition of amide in the C-terminal one. 
The total amount of peptides is quite low, and they are not evenly distributed. This is due to the fact that they are the combinatorial combination of the 20 amino acids that exist in nature.
The dataset with the dipeptides is very small (400 ligands) while the others are bigger (8000 tripeptides and 160000 tetrapeptides).

\begin{figure}
\centering
\includegraphics[width=\columnwidth]{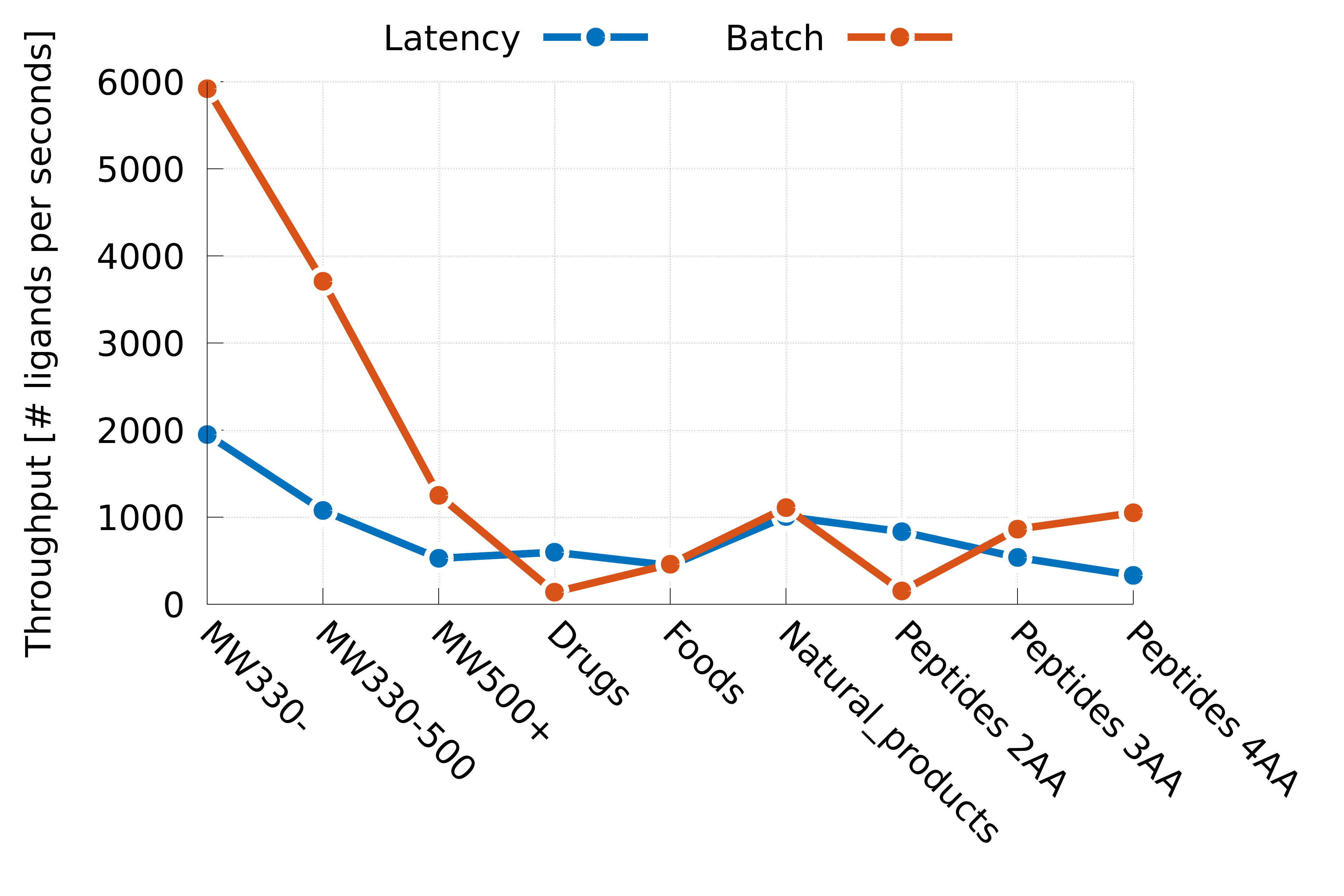}
\caption{Throughput comparison on the Mediate dataset.}
\label{fig:mediate}
\end{figure}

\prettyref{fig:mediate} reports the throughput of the different implementations on the several files composing the mediate dataset.
We can immediately notice that on the largest files (the Commercial with the different molecular weight) the batched version strongly outperforms the latency implementation. This is expected since we have 5 Million molecules here, and this amount heavily exceeds what we have found to be the cross-over point (\prettyref{ssec:scaling}).
However, the remaining files are not as big.
There are in particular two datasets, Drugs and Peptides2AA, where the batched version is unable not only to reach its optimal performances but also to reach a throughput good enough to be better than the latency implementation.
The first dataset has 14K ligands, 
%This happens in the first case because the dataset is very unbalanced: this dataset has 14K ligands,
which from the scaling analysis should be enough to at least exceed the performance of the latency implementation. However, it is unable to reach a good throughput because it is heavily unbalanced, thus in the runtime, it forces the execution of several almost empty batches which is detrimental to the performances.
On the other hand, the Peptides\_2AA is a very small dataset and, even if it is quite uniform, it still has not enough data to outperform the latency implementation.
In all the remaining datasets, the batched implementation performs closely or better than the latency but is unable to reach its peak performance. 

\begin{figure}
\centering
\includegraphics[width=\columnwidth]{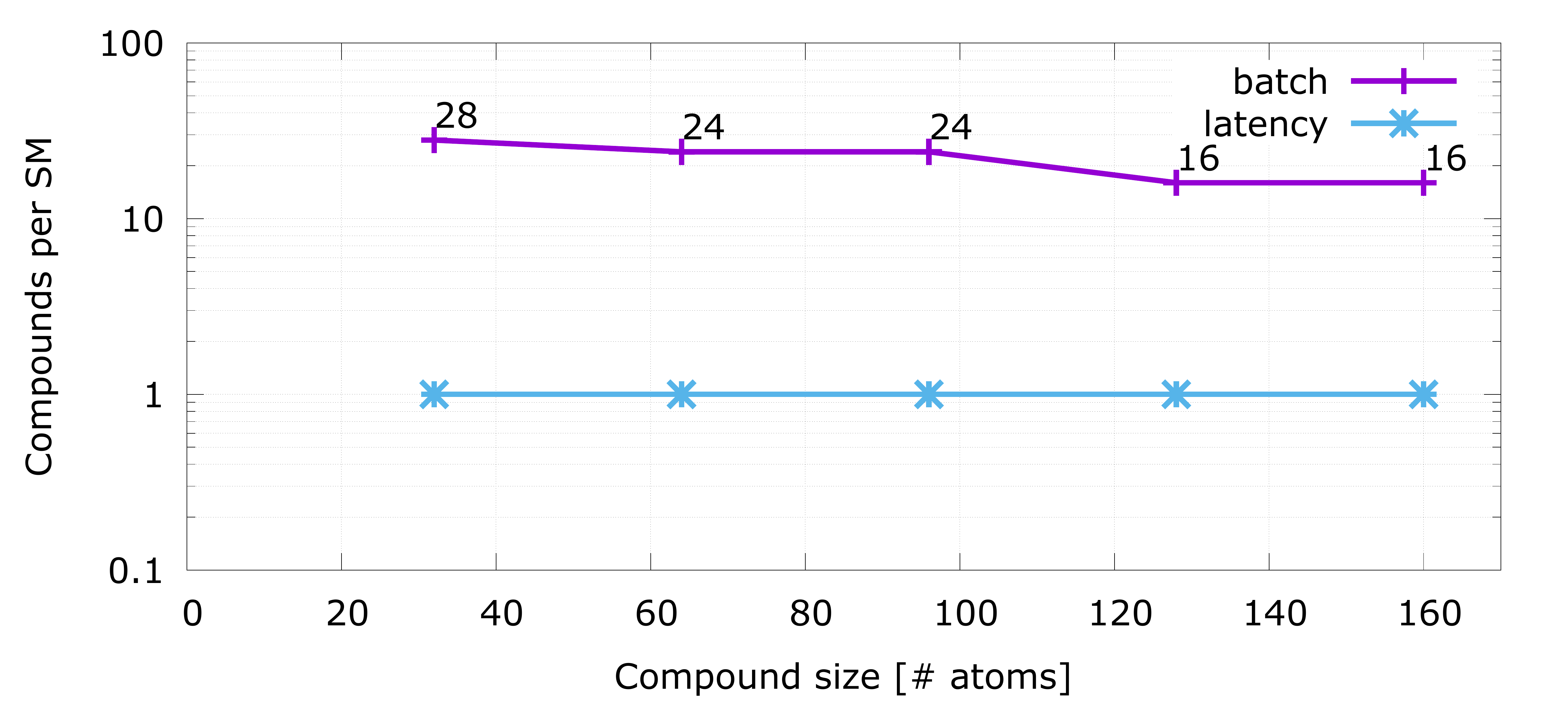}
\caption{Static ligands allocation per SM}
\label{fig:static}
\end{figure}

\begin{figure*}[t]
%	\centering
%	\begin{subfigure}{1.00\columnwidth}
%		\includegraphics[width=\columnwidth]{img/roof_lat_fp.pdf}
%		\caption{Floating point roofline for \textit{latency}}
%		\label{fig:latency_roofline_fp}
%	\end{subfigure}
%	\begin{subfigure}{1.00\columnwidth}
%		\includegraphics[width=\columnwidth]{img/roof_bat_fp.pdf}
%		\caption{Floating point roofline for \textit{batch}}
%		\label{fig:batch_roofline_fp}
%	\end{subfigure}
	
	\centering
	\begin{subfigure}{1.00\columnwidth}
		\includegraphics[width=\columnwidth]{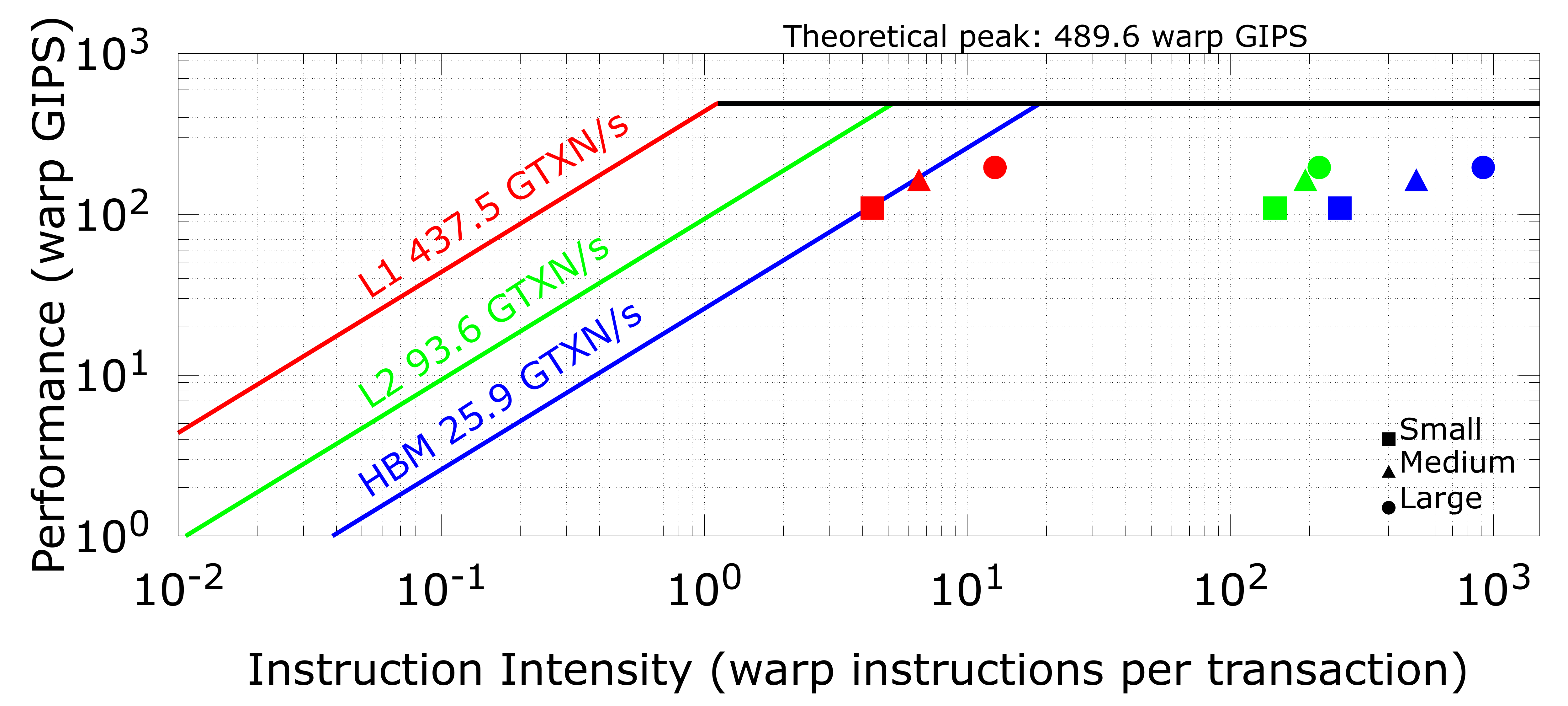}
		\caption{Instruction roofline for \textit{latency}}
		\label{fig:latency_roofline_inst}
	\end{subfigure}
	\begin{subfigure}{1.00\columnwidth}
		\includegraphics[width=\columnwidth]{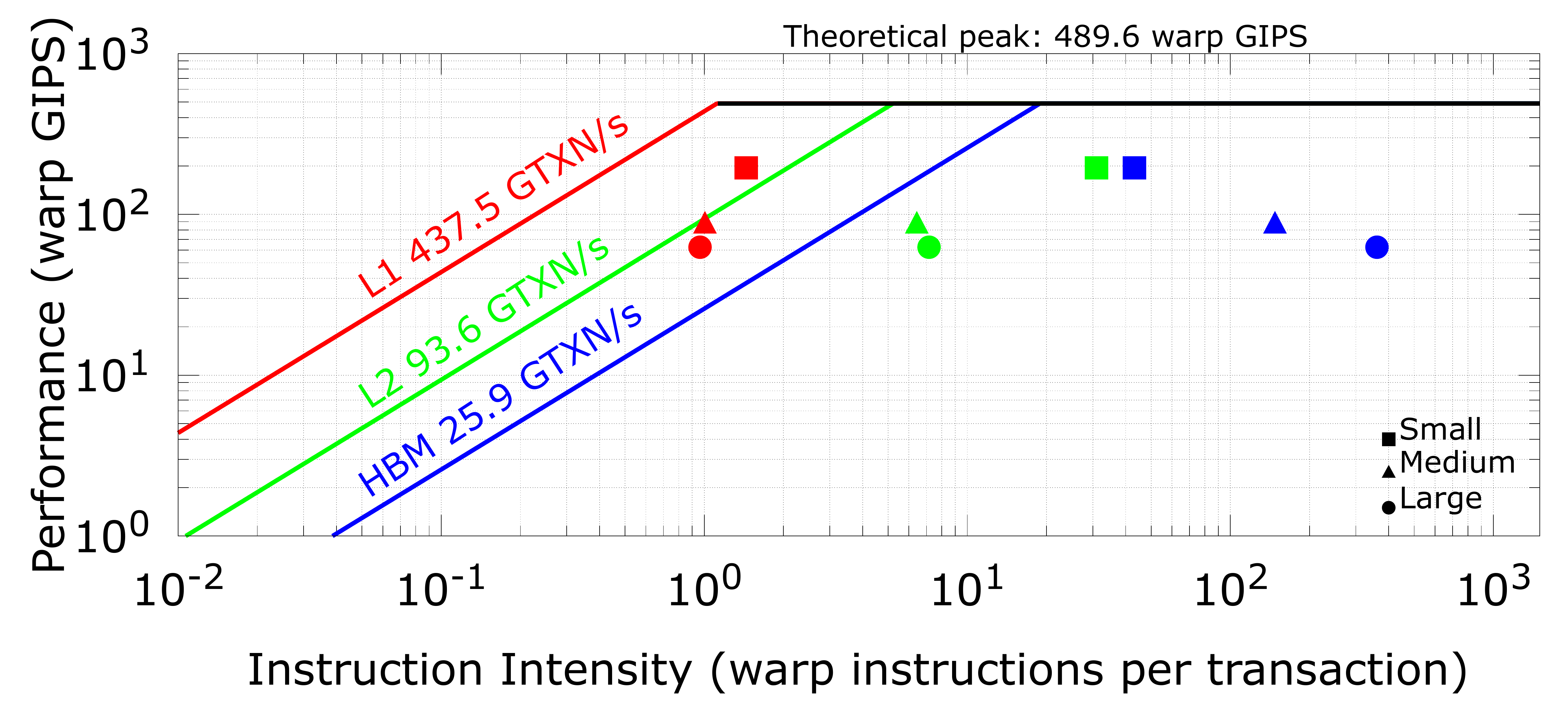}
		\caption{Instruction roofline for \textit{batch}}
		\label{fig:batch_roofline_inst}
	\end{subfigure}
	
	\centering
	\begin{subfigure}{1.00\columnwidth}
		\includegraphics[width=\columnwidth]{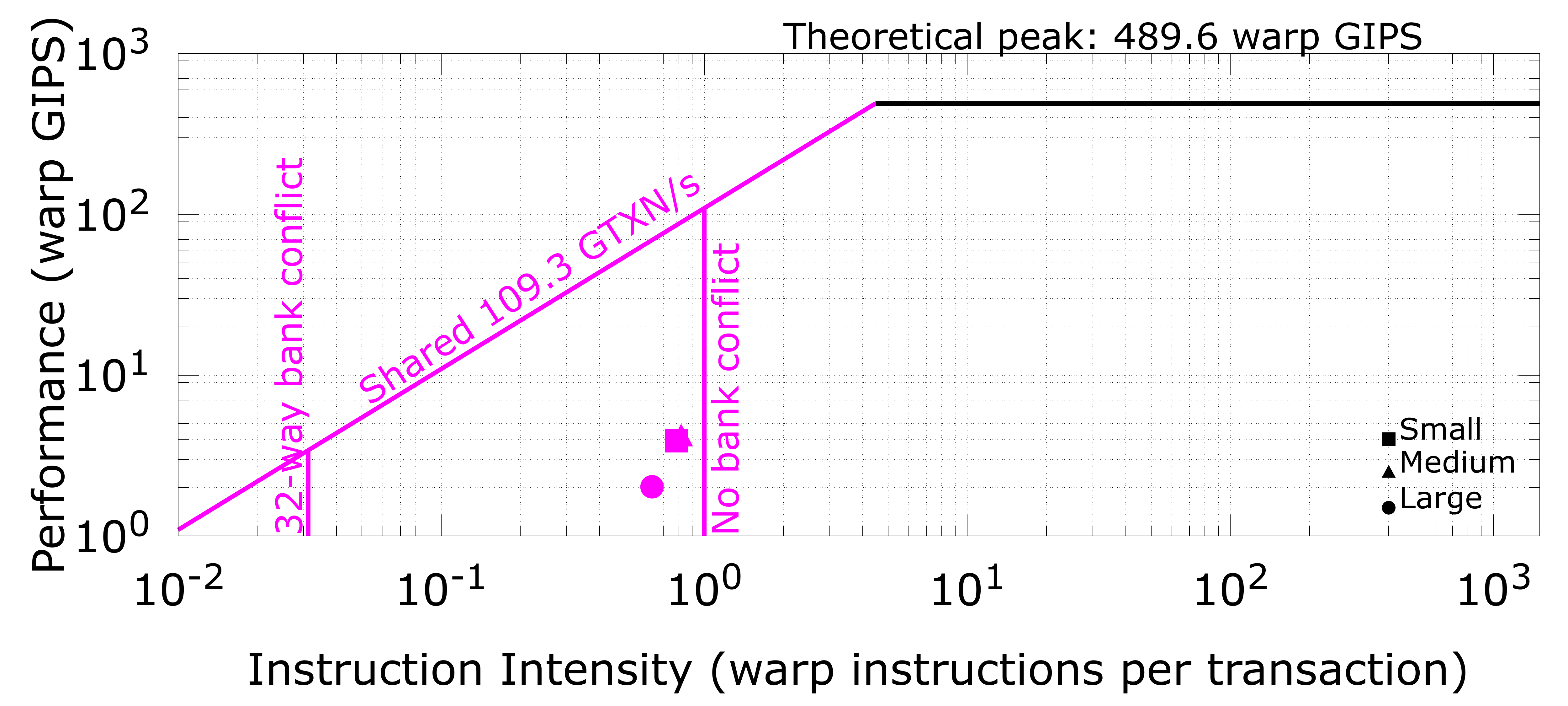}
		\caption{Shared memory access pattern roofline for \textit{latency}}
		\label{fig:latency_roofline_shared}
	\end{subfigure}
	\begin{subfigure}{1.00\columnwidth}
		\includegraphics[width=\columnwidth]{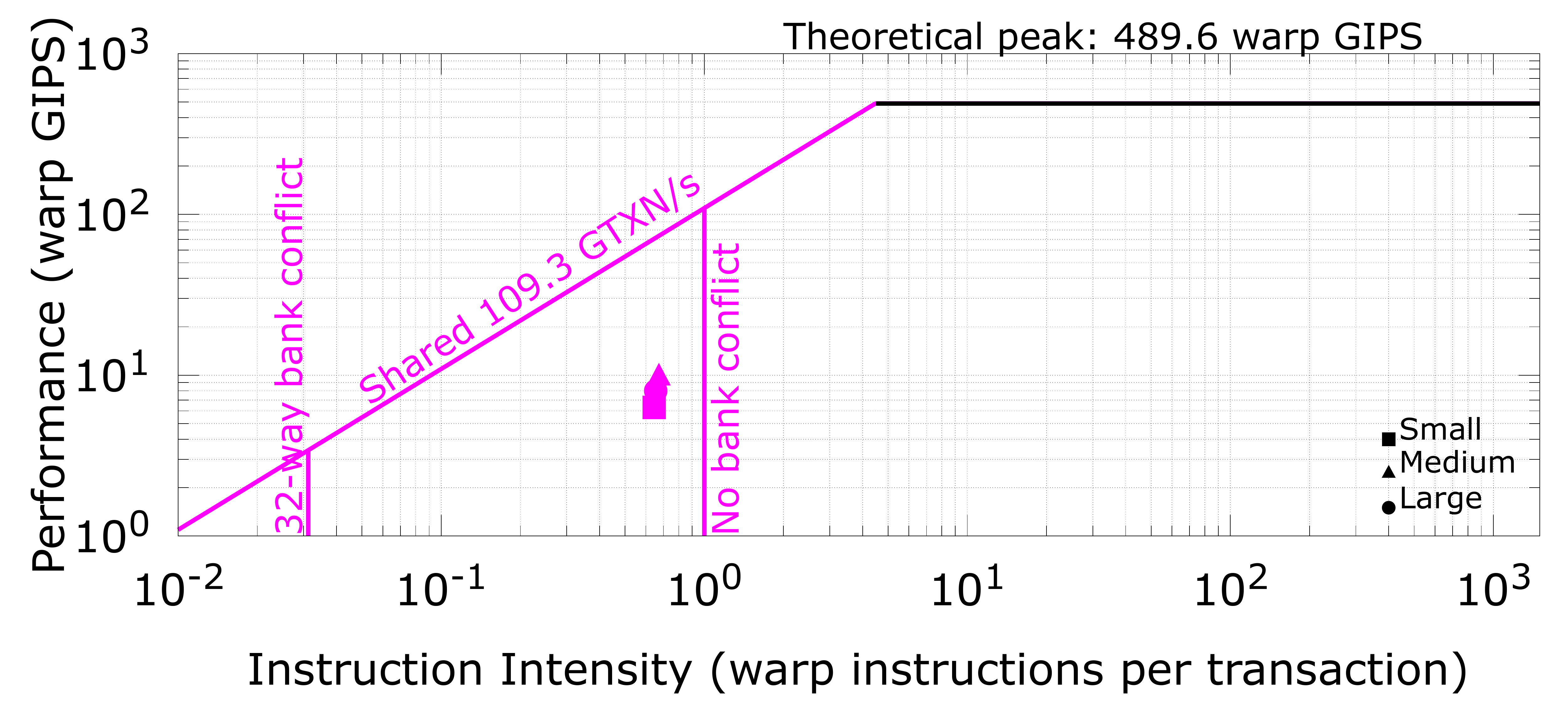}
		\caption{Shared memory access pattern roofline for \textit{batch}}
		\label{fig:batch_roofline_shared}
	\end{subfigure}
	\caption{Roofline analysis comparison between \textit{latency} (left) and \textit{batch} (right) on  instruction performance (\prettyref{fig:latency_roofline_inst} and \prettyref{fig:batch_roofline_inst}) and shared memory access pattern (\prettyref{fig:latency_roofline_shared} and \prettyref{fig:batch_roofline_shared})}
	\label{fig:roofline}
\end{figure*}

\subsection{Workload Analysis}
\label{ssec:workload}

With the previous analysis, we have seen that the batched implementation has a slow start but a better overall throughput.
Now we want to analyze more in-depth the two implementations to search for the reason behind this result.
To reach this goal, we will characterize both workloads in terms of execution profiles,
applying the \textit{instruction roofline methodology}\cite{ding2019instruction}, on an input dataset constructed to be representative of different molecule categories from real-world datasets \cite{mediate}.

Considering both dimensions that affect workload's computational complexity, namely the number of atoms (including hydrogens) and the number of rotatable bonds, three molecule classes have been defined. For each of those classes, a sample molecule from the testing dataset has been randomly selected and then duplicated to produce a uniform input batch up to the suggested size as described in \prettyref{sec:batched}: a uniform input dataset allows for homogeneous execution paths across all warps involved in a single kernel grid, especially for the batched implementation where each warp takes care of different input ligands.
%Changing the molecule It has been seen
The test molecule classes have been defined as:
\begin{itemize}
	\item \textit{Small}: \((0, 64]\) atoms, 1 rotatable bond, batch of 1920 molecules;
	\item \textit{Medium}: \((64, 96]\) atoms, 12 rotatable bonds, batch of 1600 molecules;
	\item \textit{Large}: \((96, 160]\) atoms, 20 rotatable bonds, batch of 960 molecules.
\end{itemize}

Since we want to know why the two implementation throughput is so different, we focus our analysis on the CUDA kernel that represents the major hotspot for each implementation.
For the \textit{latency} version, this is the kernel that performs the ligand's fragment optimization (lines 10-16 in \prettyref{alg:one}, accounting for 92\% of the overall docking pipeline's runtime).
While for the \textit{batch} version, the kernel performs the full docking procedure (lines 1 - 17, accounting for 93\% of the overall docking pipeline's runtime).

%Regarding the workload taken into account for the analysis, we have considered only a single kernel for each implementation by identifying the single CUDA kernel where the whole kernel pipeline has its major hotspot: the kernel performing the ligand's fragment optimization for \textit{latency} (accounting for 92\% of the overall docking pipeline's runtime) and the kernel performing the full docking procedure for \textit{batch} (accounting for 93\% of the overall docking pipeline's runtime).

\subsubsection{Resources Allocation}
\label{sec:workload_static}

To understand the consequences of different design principles between the two approaches, we analyzed static resource allocation first.

In \prettyref{fig:static} is shown the maximum amount of ligands that are allocated on a single SM.
While the \textit{latency} version dedicates all the resources within an SM to a single ligand, the \textit{batch} version allocates multiple ligands to a single warp allowing for multiple concurrently running ligands in a single SM. 
In the latter implementation, the registers per thread are the limiting factor for the ligands allocation to an SM. Therefore, the number of ligands assigned to an SM decreases with the increment of their complexity.

%Since statically allocated resources increase according to the ligand complexity class, the amount of \textit{Large} molecules allocated to a single SM decreases compared to less complex classes.
This has two evident consequences: on the one hand, the latency implementation has a more consistent behavior that does not depend on the ligand size, while on the other hand the batched implementation is strongly influenced by the data size.
It has an optimal behavior with small ligands, and it degrades increasing the size of the ligand.

Moreover, we can see that the batched implementation is able to process more ligands per SM, and this allows it to reduce the overheads when launching the kernels since it will have a smaller amount of kernels to launch. 
Indeed, while in the latency implementation we have to launch at least 1 kernel per ligand, in the batched one we process between 960 and 1920 ligands with a single kernel.
%\textbf{GIANLUCA: POSSIAMO DARE QUALCHE CONCLUSIONE O EFFETTO SU QUESTO? - e.g. This results in a better resource usage for batch?}

%Given the different design principles adopted by the two implementations, the analysis shows that, while \textit{latency} allocates the same amount of ligands to a single SM regardless of their size (and thus, complexity), \textit{batch}, by allocating multiple ligands to a single warp allows for multiple concurrently running ligands for a single SM; since statically allocated resources increase according to the ligand complexity class (registers per thread being the limiting factor), the amount of \textit{Large} molecules allocated to a single SM decrease compared to less complex classes.

\subsubsection{Execution Profiling}
\label{sec:workload_dynamic}

%girare mettendo prima cosa voglio cercare poi immagine (i.e. entrambe le implementazioni usano bene la gpu xk nel roofline si vede che blah blah).

In this section, we analyze how the computing resources are used by the kernels, to search for the reason for the different behavior between the two versions in their throughput scaling.
We present a comparison between different roofline plots \cite{ding2019instruction} produced by measuring both implementations execution behaviours via NVIDIA NSight profiler\cite{nvidia2022nsight} in \prettyref{fig:roofline}.
%According to the \textit{floating point roofline} shown in \prettyref{fig:latency_roofline_fp} and \prettyref{fig:batch_roofline_fp}, neither implementation is bound to any specific memory or computational limit. 

In particular, in \prettyref{fig:latency_roofline_inst} and \prettyref{fig:batch_roofline_inst} we report the instruction issued roofline. These rooflines are obtained by considering all kinds of warp-level instructions issued. From these two graphs, we can say that both the implementations of the application are not memory-bound. Moreover, since we are close to the roof, we can say that in both implementations we are using the GPU appropriately.
We can notice a difference in the two implementations if we look at their behavior on the size of the different molecules. On the one hand, in the \textit{batch} implementation the amount of issued instruction per second decreases with the increase of the molecule size, while on the other hand in the \textit{latency} implementation it increases. 
This is expected due to respective scaling design choices: on the latency version, we improve the number of instructions because the efficiency of the kernel is constant, and the amount of data increases. On the other hand in the batched implementation the amount of instructions decreases because we are using more registers to store the ligands and in this way, we have fewer active threads per SM, and this decreases the number of operations executed.

\begin{figure}[t]
	\centering
	\begin{subfigure}{0.99\columnwidth}
		\includegraphics[width=\columnwidth]{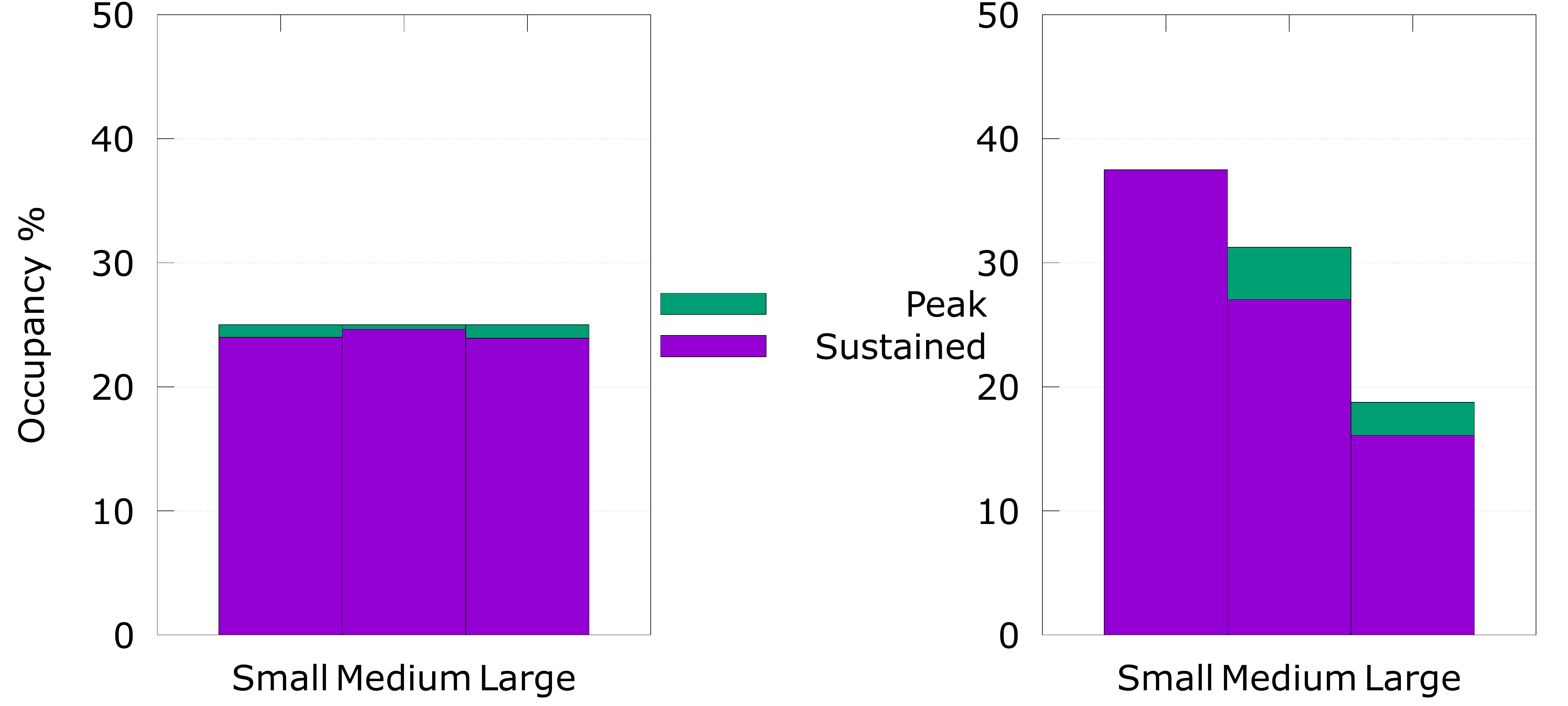}
		\caption{SM occupancy}
		\label{fig:diff_hist_occupancy}
	\end{subfigure}
	
	\begin{subfigure}{0.99\columnwidth}
		\includegraphics[width=\columnwidth]{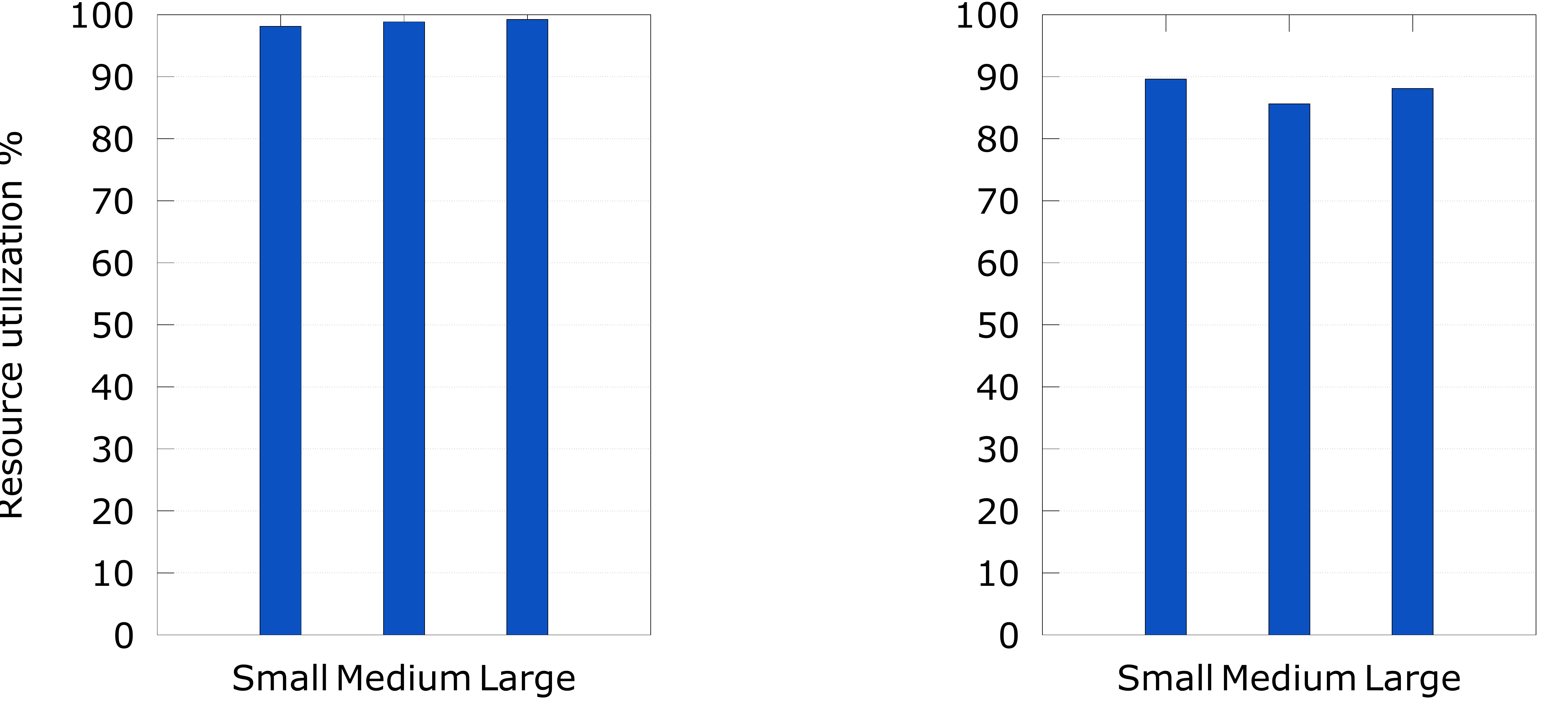}
		\caption{Efficiency (thread predication)}
		\label{fig:diff_hist_efficiency}
	\end{subfigure}
		
	%\begin{subfigure}{0.99\columnwidth}
	%	\includegraphics[width=\columnwidth]{img/performance.pdf}
	%	\caption{Floating point performance}
	%	\label{fig:diff_hist_perf}
	%\end{subfigure}
	
	\begin{subfigure}{0.99\columnwidth}
		\includegraphics[width=\columnwidth]{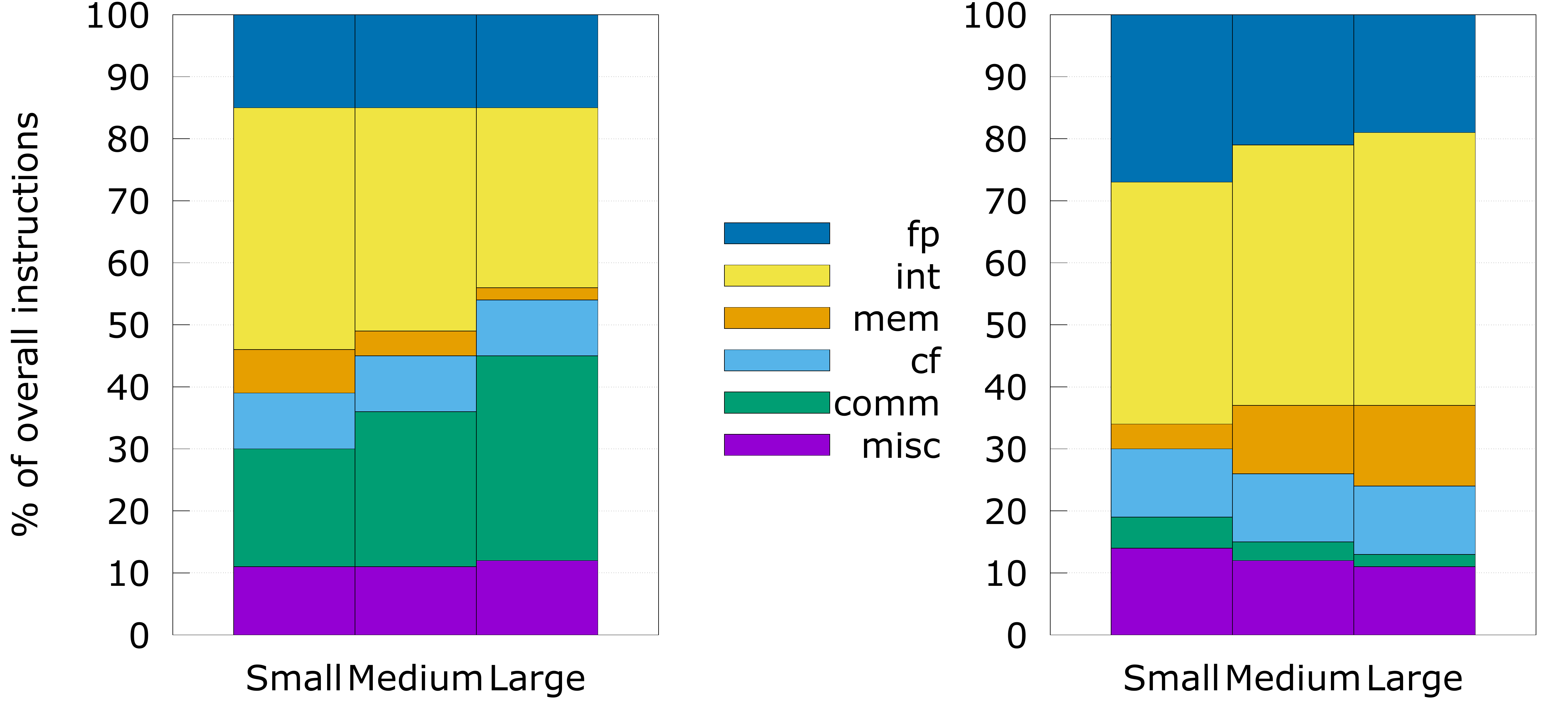}
		\caption{Instruction mix}
		\label{fig:diff_hist_instmix}
	\end{subfigure}

    \caption{Comparison between \textit{latency} (left) and \textit{batch} (right) on peak and sustained active warps (\prettyref{fig:diff_hist_occupancy}), efficiency (or thread predication, \prettyref{fig:diff_hist_efficiency}) and instruction mix (\prettyref{fig:diff_hist_instmix}).}
    %\caption{Comparison between \textit{latency} (left) and \textit{batch} (right) on peak and sustained active warps (\prettyref{fig:diff_hist_occupancy}), efficiency (or thread predication, \prettyref{fig:diff_hist_efficiency}), instruction mix (\prettyref{fig:diff_hist_instmix}) and floating point performance (\prettyref{fig:diff_hist_perf}).}
    \label{fig:diff_hist}
\end{figure}

Another insight given by these two plots is the cache reuse: 
the horizontal distance between points of the same molecule class represents the ability of the cache to satisfy a request. The larger the distance between two points, the highest the reuse of data present in the highest level memory (i.e. distance between L1 and L2 caches represents the ability of the L1 cache to serve the read request).

The \textit{latency} implementation (\prettyref{fig:latency_roofline_inst}) shows regular cache reuse across molecule classes, and we can notice that the reuse of the L2 cache increases with the size of the ligands.
On the other hand, the \textit{batched} implementation (\prettyref{fig:batch_roofline_inst}) have a high L1 reuse for \textit{Small}: L1 arithmetic and instruction intensities are $\sim10\times$ lower than L2 and HBM values.
However, larger molecule classes begin to rely heavily on L2 cache: this can be seen by the fact that the HBM arithmetic and instruction intensities are $\sim100\times$ higher than L1 and L2 values. This also strengthens the idea that the batched implementation has a better behavior with small molecules but it degrades with the growth of the data size.
%we can notice that the L1 Instruction Intensity is lower than the one obtained by small molecules, and it is decreasing. Moreover, the L2 cache is more used (shown by the HBM arithmetic and instruction intensities being $\sim100\times$ higher than L1 and L2 values) hinting at \textit{batch} design imbalance towards small molecules.

The second set of images reports the \textit{shared memory roofline} (\prettyref{fig:latency_roofline_shared} and \prettyref{fig:batch_roofline_shared}). 
They are obtained by measuring both warp-level load/store instructions issued and shared memory transactions performed).
The x-axis indicates within the interval between no bank conflict and 32-way bank conflicts how efficient the kernel is in terms of shared memory access. 
Both implementations show little to no impact due to shared memory bank conflicts and thus an efficient access pattern.

From this analysis, the two implementations seem similar, with the batched one showing a better utilization of the GPU for small ligands, while the latency one uses the resources in a better way with large ligands.
However, this analysis is unable to explain the speedups that we have found from the experiments done in \prettyref{ssec:preprocessed_ds},\prettyref{ssec:realdb} and \prettyref{ssec:scaling}.
%we know that the batched implementation has always a speedup when compared to the latency one, given enough data to process.

For this reason, we need to further investigate the execution profiles of the two algorithm implementations.
The results of this analysis are reported in \prettyref{fig:diff_hist}.
The first image reports the \textit{occupancy} (\prettyref{fig:diff_hist_occupancy}), defined as the ratio between sustained and peak percentage of active warps per SM (measured by \texttt{sm\_\_maximum\_warps\_per\_active\_cycle\_pct} and 	\texttt{sm\_\_warps\_active.avg.pct\_of\_peak\_sustained\_active} metrics respectively \cite{nvidia2022nsight}).
Occupancy is one of the factors that can be used to improve performances, but it's not the only one since it is possible to reach optimal performances by decreasing the occupancy and having more registers per thread \cite{volkov2010better}. For this reason, we are not interested in the absolute value in this graph but we are looking at the comparison between the two implementations.
Both implementations show a comparable degree of SM occupancy.
We can notice that while for \textit{batch} it decreases with an increasing molecule complexity (more registers used), for \textit{latency} the behavior is uniform.%, both according to their respective design principles.
This analysis does not provide any insight into the difference in throughput but helps in explaining why the advantage of using the batched implementation decreases with larger molecules.

The second image reports the \textit{efficiency} (\prettyref{fig:diff_hist_efficiency}), defined as the degree of thread predication across all the instructions executed in a single \textit{SM Sub Partition} (or SMSP\cite{nvidia2022nsight}, measured by the \texttt{smsp\_\_thread\_inst\_executed.sum} metric for thread-level instructions and \texttt{smsp\_\_inst\_executed.sum} for warp-level instructions). Both implementations show high degrees of execution efficiency and thus low degrees of thread predication. Slightly higher predication in \textit{batch} is to ascribe to molecule sizes not being a multiple of the warp size. 
%No interesting information can be gained from this plot, both implementations are quite efficient.
This plot demonstrates how both implementations are quite efficient in the use of resources.

% Io questa la rimuoverei
%The third image reports the \textit{floating point throughput} (\prettyref{fig:diff_hist_perf}), defined as the attained FLOP/s considering any floating point precision instructions. Again, we can see the already analyzed trend with the latency implementation showing a growing amount of operations, while the batched one starts higher to drop with the increasing complexity of the ligands.

Finally the third image reports the \textit{instruction mix} (\prettyref{fig:diff_hist_instmix}), defined as the percentage of instructions executed in a single SMSP grouped by instruction type:
\begin{itemize}
    \item \texttt{fp}: floating point instructions (any precision, including scalar, FMA and tensor),
    \item \texttt{int}: integer instructions (any integer data type),
    \item \texttt{mem}: memory operations (load/stores),
    \item \texttt{cf}: control flow operations,
    \item \texttt{comm}: inter-thread communication and synchronization,
    \item \texttt{misc}: everything else including bit-wise operations and casts
\end{itemize}
There are two interesting pieces of information in this figure.
The first one is that the largest part of the operation done is integer arithmetic. This is expected since they comprehend index calculations and the \textit{Score} function used to select the best pose is a sum over integer values.
Moreover, if we look at the latency implementation, we can notice that it has a large (20 to 40 \%) of \texttt{comm} instructions, that almost completely disappear in the batched implementation.
These \texttt{comm} instructions are mostly due to the design of the latency kernel.
In \prettyref{sec:application_descr} we have the pseudocode of the algorithm. As we already mentioned, in the pose optimization phase (lines 11-15) we need to process all the fragments sequentially. 
The most complex function in this phase is the \textit{CheckBump} (line 13) function.
Indeed, in this function, we need to analyze all the atoms of the ligand to check that they are not too close to another atom of the ligand. If we find one bump, we need to invalidate the pose.
This introduces a problem for the latency kernel: since we distribute the computation across an entire SM, we have two options. On the one hand, we can always evaluate all the atoms, then synchronize and accept or reject the pose. This would lead to a slowdown since we lose the opportunity of avoiding useless computations whenever a bump is found. On the other hand, to maintain the early exit possibility, we need to introduce a lot of synchronization points inside the kernel. This approach is still faster than doing all the atoms for every pose, but it still introduces a large overhead.
This overhead is not present at all in the batched approach where we evaluate one ligand within a single warp. This means that we don't need to synchronize the SM when doing this phase, and we can use the early exit without overheads.

\begin{figure}
\centering
\includegraphics[width=\columnwidth]{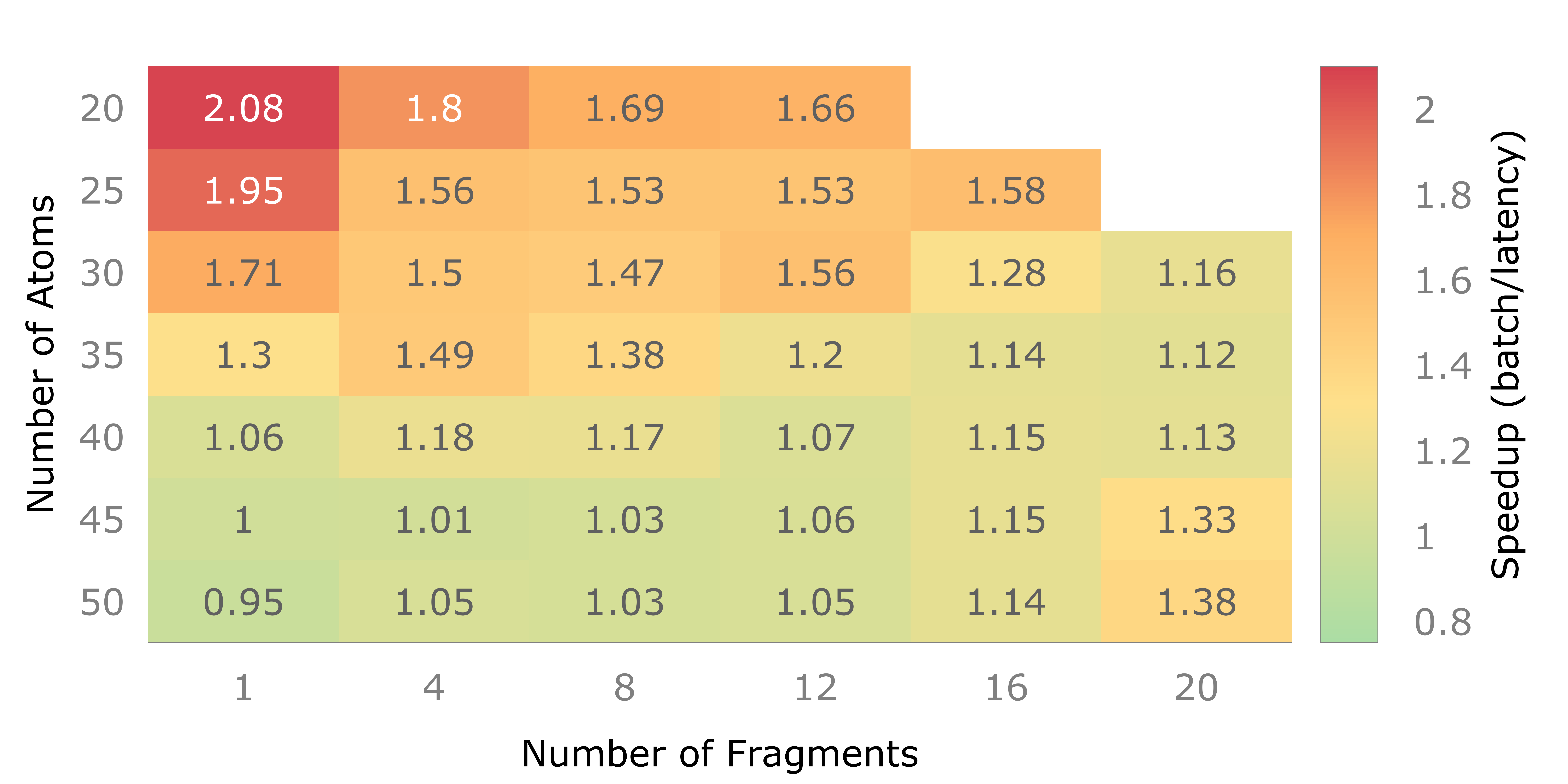}
\caption{Speedup Heatmap of the batched version against the latency one for the different homogeneous datasets without the early exit from the \textit{CheckBump} function. Both throughputs are taken with large enough datasets.}
%\textbf{number atoms total}}
\label{fig:speedup_nobump}
\end{figure}

To confirm this hypothesis, we have run both implementations without the early escape from this loop, and we report the result in \prettyref{fig:speedup_nobump}. 
We can notice that the advantage in terms of speedup in using the batched approach has been reduced a lot and reached a maximum value of 2x only for very small ligands.
This is expected since in the previous analysis we have seen that for small molecules the batched approach is more efficient.
On other molecule dimensions, i.e. larger in terms of atoms and fragments), the speedup is slightly above 1, including a small slowdown for the bottom left corner. 
%We can notice that we don't have anymore a great advantage in using the batched approach, as the speedup, in this case, has a maximum value of 2x (for small ligands, that is expected since in the previous analysis we have seen that for small molecules is more efficient). On the other side of the spectrum, for large ligands, we can notice also a slight slowdown. 
%Increasing the number of fragments, the speedup is more or less flattened around 1 and 
This analysis confirms that the management of the early exit condition is the tie-breaker between the two implementations since in the batched version we can use it without introducing a lot of synchronization overhead. 

The \textit{latency} implementation, given its design principle of scaling out computing resources according to the input ligand's complexity, shows a regular behavior across molecule classes in terms of performance, occupancy, and instruction throughput.

On the other hand, the \textit{batch} implementation uses a fixed amount of computing resources allocated to a batch of input ligands and deals with the increasing molecules' complexity by increasing the amount of work a single warp must carry out. 
Moreover, this second implementation has its best behavior with small molecules and its performances have a slight degradation when increasing the data size because fewer compute resources are used since we need more registers for the data thus decreasing the number of active threads.

To conclude this discussion, we have seen that the great advantage of using a batched approach is mainly due to the fact that by using a warp to analyze a ligand we can avoid most of the synchronization between warps in the same SM. 
This is fundamental in the \textit{CheckBump} function because it allows the exploitation of the early exit condition without introducing too much overhead.
%As shown by the analysis, this guarantees a significant advantage on \textit{Small} molecules and a decreasing advantage with increasing molecule size.

\section{Conclusion}
\label{sec:conclusion}
In this paper, we have presented the problem of virtual screening a large set of molecules. 
We have seen that it is usually tackled by performing molecular docking of the candidate molecules in the protein pocket, and this process is done by using large computer simulations.
We have presented two optimized implementations of a molecular docking application designed for virtual screening that uses the GPU as a hardware accelerator for the docking procedure. 
While the first version refers to the classical \emph{latency} approach that spreads the computation of a ligand-protein pair across the device, the second one focuses more on the throughput of a virtual screening campaign. In this second version, we process a \emph{batch} of ligand-protein pairs across the device, increasing the latency of a single evaluation but improving the throughput of the whole screening.
We compare the different ideas behind the two approaches, and thanks to an extensive experimental section, we compare the two implementations to search for their limits and advantages.

% if have a single appendix:
%\appendix[Proof of the Zonklar Equations]
% or
%\appendix  % for no appendix heading
% do not use \section anymore after \appendix, only \section*
% is possibly needed

% use appendices with more than one appendix
% then use \section to start each appendix
% you must declare a \section before using any
% \subsection or using \label (\appendices by itself
% starts a section numbered zero.)

 %\appendix[Metrics for data collection]
 %\input{appendix}

% use section* for acknowledgment
\ifCLASSOPTIONcompsoc
  % The Computer Society usually uses the plural form
  \section*{Acknowledgments}
\else
  % regular IEEE prefers the singular form
  \section*{Acknowledgment}
\fi
This work has received funding from EuroHPC-JU under the grant agreement No 956137 (LIGATE), and from H2020 Programme under the grant agreement No 101003551 (Exscalate4CoV).

%The authors would like to thank...

% Can use something like this to put references on a page
% by themselves when using endfloat and the captionsoff option.
\ifCLASSOPTIONcaptionsoff
  \newpage
\fi

% trigger a \newpage just before the given reference
% number - used to balance the columns on the last page
% adjust value as needed - may need to be readjusted if
% the document is modified later
%\IEEEtriggeratref{8}
% The "triggered" command can be changed if desired:
%\IEEEtriggercmd{\enlargethispage{-5in}}

% references section

% can use a bibliography generated by BibTeX as a .bbl file
% BibTeX documentation can be easily obtained at:
% http://mirror.ctan.org/biblio/bibtex/contrib/doc/
% The IEEEtran BibTeX style support page is at:
% http://www.michaelshell.org/tex/ieeetran/bibtex/
%\bibliographystyle{IEEEtran}
% argument is your BibTeX string definitions and bibliography database(s)
%\bibliography{IEEEabrv,../bib/paper}
%
% <OR> manually copy in the resultant .bbl file
% set second argument of \begin to the number of references
% (used to reserve space for the reference number labels box)
\bibliographystyle{IEEEtran}
\bibliography{biblio}

% biography section
% 
% If you have an EPS/PDF photo (graphicx package needed) extra braces are
% needed around the contents of the optional argument to biography to prevent
% the LaTeX parser from getting confused when it sees the complicated
% \includegraphics command within an optional argument. (You could create
% your own custom macro containing the \includegraphics command to make things
% simpler here.)
%\begin{IEEEbiography}[{\includegraphics[width=1in,height=1.25in,clip,keepaspectratio]{mshell}}]{Michael Shell}
% or if you just want to reserve a space for a photo:

% You can push biographies down or up by placing
% a \vfill before or after them. The appropriate
% use of \vfill depends on what kind of text is
% on the last page and whether or not the columns
% are being equalized.

%\vfill

% Can be used to pull up biographies so that the bottom of the last one
% is flush with the other column.
%\enlargethispage{-5in}

% that's all folks
\end{document}